\documentclass{amsart}
\usepackage{amssymb}
\usepackage{latexsym}
\usepackage{xy}
\input xypic
\newtheorem{theorem}{Theorem}

\newtheorem{lemma}[theorem]{Lemma}

\def\Proof{\medskip\noindent{\bf Proof: }}

\def\Z{\mathbb{Z}}

\def\P{\mathbb{P}}

\def\C{\mathbb{C}}

\def\Q{\mathbb{Q}}

\def\R{\mathbb{R}}
\def\C{\mathbb{C}}

\def\Pi{\mathbb{P}^{\infty}}

\def\qed{\hfill$\square$\medskip}

\def\Zpk{\mathbb{Z}/p^{k}}
\def\Zpk1{\mathbb{Z}/p^{k-1}}

\newcommand{\rref}[1]{(\ref{#1})}

\newcommand{\cform}[3]{\begin{array}{c}
{\scriptstyle #3}\\
#1\\
{\scriptstyle #2}\end{array}}

\newcommand{\beg}[2]{\begin{equation}\label{#1}#2\end{equation}}
\def\r{\rightarrow}

\def\mh{\mathcal{H}}

\def\sl2{\widetilde{SL_{2}(\Z)}}

\author{Igor Kriz}
\title{Perturbative deformations of conformal field theories revisited}

\begin{document}

\maketitle

\section{Introduction}

Recently, there has been renewed interest in the mathematics of
the moduli space of conformal field theories, in particular
in connection with speculations about elliptic cohomology. The purpose
of this paper is to investigate this space by 
perturbative methods from first
principles and from a purely ``worldsheet'' point of view.
It is conjectured
that at least at generic points, the moduli space of CFT's is
a manifold, and in fact, its tangent space consists of
marginal fields, i.e. primary fields of weight $(1,1)$ of the
conformal field theory (that is in the bosonic case, in the supersymmetric
case there are modifications which we will discuss later).
This then means that there should exist an
exponential map from the tangent space at a point to the moduli
space, i.e. it should be possible to construct a continuous $1$-parameter
set of conformal field theories by ``turning on'' a given 
marginal field. 

\vspace{3mm}
There is a more or less canonical mathematical procedure for applying
a ``$Pexp$'' type construction to the field which has been turned on,
and obtaining a perturbative expansion in the deformation parameter. 
This process, however, returns certain cohomological obstructions, similar
to Gerstenhaber's obstructions to the existence 
of deformations of associative algebras
\cite{gerst}. Physically, these obstructions can
be interpreted as changes of dimension of the
deforming field, and can occur, in principle,
at any order of the perturbative path. The primary obstruction is
well known, and was used e.g. by Ginsparg in his work on $c=1$
conformal field theories \cite{ginsparg}. The obstruction also
occured in earlier work, see 
\cite{kad,kadb,kadw,wilson,wilson1,wilson2,wegner}, 
from the point of view
of continuous lines in the space of critical models.
In the models considered, notably the Baxter model \cite{bax},
the Ashkin-Teller model \cite{ashten} and the Gaussian model
\cite{kohmoto}, vanishing of the primary obstruction did 
correspond to a continuous line of deformations, and it was
therefore believed that the primary obstruction tells the whole
story. (A similar story also
occurs in the case of deformations of boundary sectors, see
\cite{konref1,konref2,konref3,konref4,konref5,konref6,konref7, kondo}.)

\vspace{3mm}
In a certain sense, the main point of the present paper is
analyzing, or giving examples of, the role of the hihger obstructions.
We shall see that these obstructions can be non-zero
in cases where the deformation is believed to exist,
most notably in the case of deforming
the Gepner model of the Fermat quintic along a cc field,
cf. \cite{ag, ns, fried, vw, w1, w2, w3, 39a, 39b, 39c}. 
Some discussion of marginality of primary field in $N=2$-supersymmetric
theories to higher order exists in the literature.
Notably, Dixon, \cite{dixon} verified the
vanishing for any $N=(2,2)$-theory, and any linear combination
of cc,ac,ca and ac field, of an amplitude integral which
physically expresses the change of central charge
(a similar calculation is also given in Distler-Greene \cite{distler}).
Earlier work of Zamolodchikov \cite{za,zb} showed
that the renormalization $\beta$-function vanishes for theories
where $c$ does not change during the renormalization process. 
However, we find that the calculation \cite{dixon}
does not guarantee
that the primary field would remain marginal along the perturbative
deformation path, due to subtleties involving
singularities of the integral. The obstruction we discuss in this
paper is an amplitude integral which physically expresses
directly the change of dimension of the 
deforming field, and it turns out this may not vanish. 
We will return to this discussion in Section
\ref{sexp} below.

\vspace{3mm}
This puzzle of having obstructions where none should
appear will not be fully explained in this paper, although
a likely interpretation of the result will be
discussed. Very likely, our effect does not 
impact the general question of the
existence of the non-linear $\sigma$-model, which
is widely believed to exist
(e.g. \cite{ag, ns, fried, vw, w1, w2, w3, 39a, 39b, 39c}), but
simply concerns questions of its perturbative construction. 
One caveat is that the case we investigate here
is still not truly physical,
since we specialize to the case of $cc$ fields, which are not
real. The actual physical deformations of CFT's should occur
along real fields, e.g. a combination of a cc field and its
complex-conjugate $aa$ field (we give a discussion of this in 
case of the free field theory at the end of Section \ref{sfree}).
The case of the corresponding real field in the Gepner
model is much more difficult
to analyze, in particular it requires regularization of the
deforming parameter, and is not discussed here. 
Nevertheless,
it is still surprising that an obstruction occurs for a single
$cc$ field; for example, this does not happen in the case
of the (compactified or uncompactified) free field theory. 

\vspace{3mm}
Since an $n$'th order obstruction indeed means that
the marginal field gets deformed into a field of non-zero weight,
which changes to the order of the $n$'th power of the
deformation parameter,
usually 
\cite{ginsparg,kad,kadb,kadw,wilson,wilson1,wilson2,wegner},
when obstructions occur, one therefore concludes that the CFT
does not possess continuous deformations in the given direction. 
Other interpretations are possible.
One thing to observe is that our conclusion is only valid for
{\em purely perturbative} theories where we assume that all fields have
power series expansion in the deformation parameters with coefficients
which are fields in the original theory. This is
not the only possible scenario. Therefore, as remarked above,
our results merely indicate that in the case when our
algebraic obstruction is non-zero, non-perturbative corrections
must be made to the theory to maintain the presence of marginal fields
along the deformation path. 

\vspace{3mm}
In fact, evidence in favor of this interpretation exists in
the form of the analysis of Nemeschansky and Sen
\cite{ns,gvz} of higher order corrections to the $\beta$-function
of the non-linear $\sigma$-model. Grisaru,
Van de Ven and Zanon \cite{gvz} found that the four-loop contribution
to the $\beta$-function of the non-linear $\sigma$ model for
Calabi-Yau manifolds is non-zero, and \cite{ns} found a recipe
how to cancel this singularity by deforming the manifold 
to metric which is non-Ricci flat at higher orders of the
deformation parameter. The expansion \cite{af} used in this
analysis is around the $0$ curvature tensor, but assuming for
the moment that a similar phenomenon occurs if we expanded around
the Fermat quintic vacuum, then there are no fields present in
the Gepner model which would correspond perturbatively
to these higher order
corrections in the direction of non-Ricci flat metric: bosonically,
such fields
would have to have critical conformal dimension classically, since
the $\sigma$-model Lagrangian is classically conformally invariant 
for non-Ricci flat target K\"{a}hler manifolds. However, quantum
mechanically, there is a one-loop correction proportional to
the Ricci tensor, thus indicating that fields expressing
such perturbative deformations
would have to be of generalized weight (cf. \cite{zhang,zhang1,huang,huang1}).
Fields of generalized weight, however, are not present in the
Gepner model, which is a rational CFT, and more generally
are excluded by unitarity (see discussions in Remarks
after Theorems \ref{t1}, \ref{t2} in Section \ref{sexp}
below). Thus, although this argument is not completely
mathematical, renormalization analysis
seems to confirm our finding that
deformations of the Fermat quintic model must in general
be non-perturbative. It is also noteworthy that the $\beta$-function
is known to vanish to all orders for $K3$-surfaces because
of $N=(4,4)$ supersymmetry. Accordingly, we also find that
the phenomenon we see for the Fermat quintic is not 
present in the case of the Fermat quartic (see Section \ref{sk3} below).
It is also worth noting that other non-perturbative phenomena
such as instanton corrections also arise when passing from
$K3$-surfaces to Calabi-Yau $3$-folds (\cite{39a,39b,39c}). 
Finally, one must also remark that the proof of \cite{ns} of 
the $\beta$-function cancellation is not mathematically
complete because of convergence questions, and thus 
one still cannot exclude even 
the scenario that not all non-linear $\sigma$ models
would exist as exact CFT's, thus creating some type of
``string landscape'' picture also in this context (cf. \cite{dt}).

\vspace{3mm}
In this paper, we shall be mostly interested in the strictly perturbative
picture. 
The main point of this paper is an analysis of the algebraic obstructions
in certain canonical cases.
We discuss two main kinds of examples, namely the free field theory (both
bosonic and $N=1$-supersymmetric), and the Gepner models of the
Fermat quintic and quartic, 
which are exactly solvable $N=2$-supersymmetric conformal
field theories conjectured to be the non-linear $\sigma$-models
of the Fermat quintic Calabi-Yau $3$-fold and the Fermat quartic
$K3$-surface.
In the case of the free field theory, what happens is essentially that
all non-trivial gravitational deformations of the free field theory are
algebraically obstructed. In the case of a free theory compactified on a torus,
the only gravitational deformations which are algebraically unobstructed
come from linear change of metric on the torus. (We will focus on
gravitational deformations; there are other examples, for example the
sine-Gordon interaction \cite{sineg,cgep}, 
which are not discussed in detail here.)

\vspace{3mm}
The Gepner case deserves special attention. 
From the moduli space of Calabi-Yau $3$-folds,
there is supposed to be a $\sigma$-model map into the moduli space of CFT's. In fact, 
when we have an exactly solvable Calabi-Yau $\sigma$-model, one
gets operators in CFT corresponding to
the cohomology groups $H^{11}$ and $H^{21}$, which measure deformations
of complex structure and K\"{a}hler metric, respectively,
and these in turn give rise to {\em infinitesimal} deformations. 
Now the Fermat quintic
\beg{eintro+}{x^5+y^5+z^5+t^5+u^5=0
}
in $\C P^4$ has a model conjectured by Gepner \cite{gepner, gepner1} 
which is embedded in the tensor product of $5$ copies of the $N=2$-supersymmetric
minimal model of central charge $9/5$. The weight $(1/2,1/2)$ $cc$ and $ac$
fields correspond to the $100$ infinitesimal deformation of complex
structure and $1$ infinitesimal deformation of K\"{a}hler metric of 
the quintic \rref{eintro+}. Despite the numerical matches
in dimension, however, it is not quite correct to say that
the gravitational deformations, corresponding to 
the moduli space of Calabi-Yau manifolds, occurs by turning on
$cc$ and $ac$ fields. This is because, to preserve
unitarity, a physical deformation can
only occur when we turn on a real field, and the fields in
question are not real. In fact, the complex conjugate of a $cc$
field is an $aa$ field, and the complex conjugate of an
$ac$ field is a $ca$ field. The complex conjugate must be
added to get a real field, and a physical deformation (we 
discuss this calculationally in the case of the free field
theory in Section \ref{sfree}).

\vspace{3mm}
In this paper, we do not discuss deformations of the Gepner model
by turning
on real fields. As shown in the case of the free field
theory in Section \ref{sfree}, such deformations
require for example regularization of the deformation
parameter, and are much more difficult
to calculate. Because of this, we work
with only with the case of one $cc$ and one $ac$ field.
We will show that 
at least one $cc$ deformation,
whose real version 
corresponds to the quintics
\beg{eintro++}{x^5+y^5+z^5+t^5+u^5+\lambda x^3y^2=0
}
for small (but not infinitesimal) $\lambda$ is algebraically obstructed.
(One suspects that similar algebraic obstructions also occur for 
other fields, but the computation is too difficult at the moment; for the
$cc$ field corresponding to $xyztu$, there is some evidence suggesting
that the deformation may exponentiate.)

It is an interesting question if non-linear
$\sigma$-models of Calabi-Yau $3$-folds must also contain non-perturbative
terms. If so, likely, this phenomenon is generic, which could be a
reason why mathematicians so far discovered so few of these
conformal field theories, despite ample physical evidence of
their existence \cite{ag, ns, fried, vw, w1, w2, w3}.

\vspace{3mm}
Originally prompted by a question of Igor Frenkel, we also 
consider the case of the Fermat quartic $K3$ surface
$$x^4+y^4+z^4+t^4=0$$
in $\C P^3$. This is done in Section \ref{sk3}. It is
interesting that the problems of the Fermat quintic do not
arise in this case, and all the infinitesimally critical
fields exponentiate in the purely perturbative sense.
This dovetails with the result of Alvarez-Gaume and Ginsparg
\cite{agg} that the $\beta$-function vanishes to all orders
for critical perturbative models with $N=(4,4)$ supersymmetry,
and hence from the renormalization point
of view, the non-linear $\sigma$ model is conformal for
the Ricci flat metric on $K3$-surfaces. There are also certain
differences between the ways mathematical considerations of
moduli space and mirror symmetry vary in the $K3$ and
Calabi-Yau $3$-fold cases, which could be related to
the behavior of the non-perturbative effects. This will
be discussed in Section \ref{scd}.

\vspace{3mm}
To relate more precisely in what setup these results occur,
we need to describe what kind of deformations we are considering.
It is well known that one can obtain infinitesimal deformations
from primary fields. In the bosonic case, the weight of these
fields must be $(1,1)$, in the $N=1$-supersymmetric case in
the NS-NS sector the critical weight is $(1/2,1/2)$ and in the
$N=2$-supersymmetric case the infinitesimal deformations
we consider are along so called ac or cc fields of weight
$(1/2,1/2)$. For more specific discussion, see section \ref{sinf}
below.
There may exist infinitesimal
deformations which are not related to primary fields (see the remarks
at the end of Section \ref{sexp}).  
However, they are excluded under a certain continuity assumption
which we also state in section \ref{sinf}.

\vspace{3mm}
Therefore, the approach we follow is exponentiating infinitesimal
deformations along primary fields of appropriate weights. 
In the ``algebraic'' approach, we assume that both the primary field
and amplitudes can be updated at all points of the deformation parameter. 
Additionally, we assume one can obtain
a perturbative power series expansion in the deformation parameter,
and we do not allow counterterms of generalized weight or non-perturbative
corrections.
We describe a cohomological obstruction theory similar to Gerstenhaber's
theory \cite{gerst} for associative algebras, which in principle
controls the coefficients at individual powers of the deformation
parameter. Obstructions
can be written down explicitly under certain
conditions. This is done in 
section \ref{sexp}. The primary obstruction in fact
is the one which occurs for the
deformations of the free field theory at gravitational fields of non-zero
momentum (``gravitational waves''). In the case of the Gepner
model of the Fermat quintic, the primary obstruction vanishes
but in the case \rref{eintro++}, one can show there is an
algebraic obstruction of order $5$ (i.e given by a $7$ point
function in the Gepner model). 

\vspace{3mm}
It should be pointed out that even in the ``algebraic'' case,
there are substantial complications we must deal with.
The moduli space of CFT's is
not yet well defined. There are different definitions of conformal
field theory, for example the Segal approach \cite{scft,hk,hkd}
is quite substantially different from the vertex operator
approach (see \cite{huang} and references therein). Since these
definitions are not known to be equivalent, and their realizations
are supposed to be points of the moduli space, the space itself 
therefore cannot be defined until a particular definition is
selected. 
Next, it remains to be specified what structure there should be
on the moduli space. Presumably, there should at least be a
topology, so than we need to ask what is a nearby conformal field
theory. That, too, has not been answered.

\vspace{3mm}
These foundational questions are enormously difficult, mostly 
from the philosophical point of view: it is very easy to define
ad hoc notions which immediately turn out insufficiently general
to be desirable. Because of that, 
we only make minimal definitions
needed to examine the existing paradigm in the context outlined.
Let us, then, confine ourselves to observing that 
even in the perturbative case, the situation is not purely algebraic,
and rather involves infinite sums which need to be discussed in 
terms of analysis. For example,
the obstructions may in fact be undefined, because they may
involve infinite sums which do not converge. Such phenomenon
must be treated carefully, since it doesn't mean automatically
that perturbative exponentiation fails. In fact, because the deformed 
primary fields are only determined up to a scalar factor, there
is a possibility of regularization along the deformation
parameter. We briefly discuss this theoretically in section
\ref{sexp}, and then give an example in the case of the free
field theory in section \ref{sfree}.

\vspace{3mm}
We also briefly discuss sufficient conditions for exponentiation. 
The main method we use is the case when Theorem \ref{l1} gives
a truly local formula for the infinitesimal amplitude
changes, which could
be interpreted as an ``infinitesimal isomorphism''
in a special case. We then
give in section 
\ref{sexp} conditions under which such infinitesimal isomorphisms
can be exponentiated. This includes the case of a coset theory,
which doesn't require regularization, and a more general case
when regularization may occur. 

\vspace{3mm}
In the final sections \ref{s1}, \ref{sexam}, namely the 
case of the Gepner model, the main
problem is finding a setup for the vertex operators which would be explicit
enough to allow evaluating the obstructions in question; the positive result
is obtained using a generalisation
of the coset construction. The formulas required are obtained
from the Coulomb gas approach (=Feigin-Fuchs realization), 
which is taken from \cite{gh}.

\vspace{3mm}
The present paper is organized as follows: In section \ref{sinf},
we give the general setup in which we work, show under which condition
we can restrict ourselves to deformations along a primary field,
and derive the formula for infinitesimally deformed amplitudes, given
in Theorem \ref{l1}. In section \ref{sexp}, we discuss exponentiation
theoretically, in terms of obstruction theory, explicit formulas
for the primary and higher obstructions, and regularization.
We also discuss supersymmetry, and in the end show a mechanism
by which non-perturbative deformations may still be possible
when algebraic obstructions occur. In section \ref{sfree}, we give the
example of the free field theories, the trivial deformations which
come from $0$ momentum gravitational deforming fields, and the primary obstruction
to deforming along primary fields of nonzero momentum. In section
\ref{s1}, we will discuss the Gepner model of the
Fermat quintic, and in section \ref{sexam}, we will discuss examples
of non-zero algebraic obstructions to perturbative deformations
in this case, as well as speculations about unobstructed deformations.
In Section \ref{sk3}, we will discuss the (unobstructed) deformations
for the Fermat quartic $K3$ surface, and in Section \ref{scd}, we
attempt to summarize and discuss our possible conclusions.

\vspace{3mm}
\noindent
{\bf Acknowledgements:}
The author thanks D.Burns, I.Dolgachev, I.Frenkel, Doron Gepner,
Y.Z.Huang, I.Melnikov, K.Wendland and E.Witten for explanations
and discussions.
Special thanks to H.Xing, who contributed many useful ideas to this project
before changing his field of interest.

\vspace{10mm}

\section{Infinitesimal deformations of conformal field theories}

\label{sinf}

In a bosonic (=non-supersymmetric) CFT $\mathcal{H}$,
if we have a primary field $u$ of weight $(1,1)$, then,
as observed in \cite{scft}, we
can make an infinitesimal deformation of $\mathcal{H}$ as follows:
For a worldsheet $\Sigma$ with vacuum $U_{\Sigma}$
(the worldsheet vacuum is the same thing as the ``spacetime'', or string,
amplitude), the infinitesimal
deformation of the vacuum is
\beg{einf1}{V_{\Sigma}=\int_{x\in\Sigma} U_{\Sigma^{x}_{u}}.
}
Here $U_{\Sigma^{x}_{u}}$ is obtained by choosing a holomorphic
embedding $f:D\r\Sigma$, $f(0)=x$, where $D$ is the standard disk.
Let $\Sigma^{\prime}$ be the worldsheet obtained by cutting $f(D)$ out
of $\Sigma$, and let $U_{\Sigma^{x}_{u}}$ be obtained by gluing the
vacuum $U_{\Sigma^{\prime}}$ with the field $u$ inserted at $f(\partial D)$.
The element $U_{\Sigma^{x}_{u}}$ is proportional to $||f^{\prime}(0)||^2$,
since $u$ is $(1,1)$-primary, so it transforms the same way as a measure
and we can define the integral \rref{einf1} without coupling with
a measure. The integral \rref{einf1} is an infinitesimal deformation
of the original CFT structure in the sense that 
$$U_{\Sigma}+V_{\Sigma}\epsilon$$
satisfies CFT gluing identities in the ring $\C[\epsilon]/\epsilon^2$.

\vspace{3mm}
The main topic of this paper is studying (in this and analogous
supersymmetric cases) the question as to when the infinitesimal
deformation \rref{einf1} can be exponentiated at least to perturbative
level, i.e. when there exist for each $n\in\mathbb{N}$ elements
$$u^0,...,u^{n-1}\in\mh, \; u^0=u$$
and for every worldsheet $\Sigma$
$$U^{0}_{\Sigma},...,U^{n}_{\Sigma}\in\bigotimes \mh^*\otimes \mh$$
such that
\beg{einf2}{U_{\Sigma}(m)=\cform{\sum}{i=0}{m} U^{i}_{\Sigma}\epsilon^i,
U^{0}_{\Sigma}=U_{\Sigma} 
}
satisfy gluing axioms in $\C[\epsilon]/\epsilon^{m+1}$,
$0\leq m\leq n$,
\beg{einf3}{u(m)=\cform{\sum}{i=0}{m}u^i\epsilon^i
}
is primary of weight $(1,1)$ with respect to \rref{einf2}, $0<m\leq n$,
and
\beg{einf4}{
\frac{dU_{\Sigma}(m)}{d\epsilon}=\int_{x\in\Sigma}U_{\Sigma^{x}_{u(m-1)}}(m-1)
}
in the same sense as in \rref{einf1}.

\vspace{3mm}
We should remark that a priori, it is not known that all deformations
of CFT come from primary fields: One could, in principle, simply ask
for the existence of vacua \rref{einf2} such that \rref{einf2} satisfy
gluing axioms over $\C[\epsilon]/\epsilon^{m+1}$. As remarked in 
\cite{scft}, it is not known whether all perturbative deformations
of CFT's are obtained from primary fields $u$ as describe above.
However, one can indeed prove that the primary fields $u$
exist given suitable {\em continuity assumptions}. Suppose the
vacua $U_{\Sigma}(m)$ exist for $0\leq m\leq n$. We notice that
the integral on the right hand side of \rref{einf4} is, by definition,
the limit of integrals over regions $R$ which are proper subsets of 
$\Sigma$ such that the measure of $\Sigma-R$ goes to $0$ (fix an analytic
metric on $\Sigma$ compatible with the complex structure). 
Let, thus, $\Sigma_{D_1,...,D_k}$ be a worldsheet obtained from
$\Sigma$ by cutting out disjoint holomorphically embedded copies
$D_1,...,D_k$ of the unit disk $D$. Then we calculate
$$\begin{array}{l}
\frac{dU_{\Sigma}(m)}{d\epsilon}=\int_{x\in\Sigma}U_{\Sigma^{x}_{u(m-1)}}(m-1)\\
=\cform{\lim}{\mu(\Sigma_{D_1,...,D_k})\r0}{}U_{\Sigma_{D_1,...,D_k}}(m-1)
\int_{x\in \bigcup D_i} U_{(\bigcup D_i)_{u(m-1)}^{x}}(m-1)\\
=\cform{\lim}{\mu(\Sigma_{D_1,...,D_k})\r0}{}\cform{\sum}{i}{}U_{\Sigma_{D_i}}(m-1)
\int_{x\in D_i} U_{(D_i)_{u(m-1)}^{x}}(m-1)\\
=\cform{\lim}{\mu(\Sigma_{D_1,...,D_k})\r0}{}\cform{\sum}{i}{}U_{\Sigma_{D_i}}(m-1)
\frac{
dU_{D_i}(m)}{d\epsilon}
\end{array}
$$
assuming \rref{einf4} for $\Sigma=D$,
so the assumption we need is
\beg{einf5}{\frac{dU_{\Sigma}}{d\epsilon}=\cform{\lim}{\mu(
\Sigma_{D_1,...,D_k})\r0}{} \cform{\sum}{i}{}U_{\Sigma_{D_i}}(m)\circ \frac{
dU_{D_i}(m)}{d\epsilon}.
}
The composition notation on the right hand side means gluing.
Granted \rref{einf5}, we can recover $\frac{d U_{\Sigma}(m)}{d\epsilon}$
from $\frac{d U_{D}(m)}{d\epsilon}$ for the unit disk $D$. Now
in the case of the unit disk, we get a candidate for $u(m-1)$
in the following way:

\vspace{3mm}
Assume that $\mh$ is topologically spanned by subspaces
$\mh_{(m_1,m_2)}$ of $\epsilon$-weight $(m_1,m_2)$ where
$m_1,m_2\geq 0$, $\mh_{(0,0)}=\langle U_D\rangle$.
Then $U_{D}(m)$ is invariant under rigid notation, so
\beg{einf6}{U_{D}(m)\in\cform{\hat{\bigotimes}}{k\geq 0}{}
\mh_{(k,k)}[\epsilon]/\epsilon^{m+1}.
}
We see that if $A_q$ is the standard annulus with
boundary components $S^1$, $qS^1$ with standard parametrizations,
then
\beg{einf7}{u(m-1)=\cform{\lim}{q\r 0}{}\frac{1}{||q||^2}U_{A_q} 
\frac{dU_{D}(m)}{d\epsilon}
}
exists and is equal to the weight $(1,1)$ summand of \rref{einf6}.
In fact, by \rref{einf5} and the definition of integral, we
already see that \rref{einf4} holds. We don't know however yet
that $u(m-1)$ is primary. To see that, however, we note that
for any annulus $A=D-D^{\prime}$ where $f:D\r D^{\prime}$
is a holomorphic embedding with derivative $r$,
\rref{einf5} also implies (for the same reason - the exhaustion
principle) that \rref{einf4} is valid with $u(m-1)$ replaced
by 
\beg{einf*}{\frac{U_{A}u(m-1)}{||r||^2}.}
Since this is true for any $\Sigma$, in particular where $\Sigma$
is any disk, the integrands must be equal, so \rref{einf*} and
$u(m-1)$ have the same vertex operators, so at least in the 
absence of null elements,
\beg{einf8}{\frac{U_{A}u(m-1)}{||r||^2}=u(m-1)
}
which means that $u(m-1)$ is primary.

\vspace{3mm}
We shall see however that there are problems with this formulation
even in the simplest possible case: Consider the free (bosonic)
CFT of dimension $1$, and the primary field $x_{-1}\tilde{x}_{-1}$.
(We disregard here the issue that $\mh$ itself lacks a satisfactory
Hilbert space structure, see \cite{hkd}, we could eliminate this
problem by compactifying the theory on a torus or by considering
the state spaces of given momentum.) Let us calculate
\beg{einf9}{\begin{array}{l}U_{D}^{1}=
\int_D \exp(zL_{-1})\exp(\overline{z}\tilde{L}_{-1})x_{-1}\tilde{x}_{-1}
\\
=\frac{1}{2}\cform{\sum}{k\geq 1}{}\frac{x_{-k}\tilde{x}_{-k}}{k}.
\end{array}
}
We see that the element \rref{einf9} is not an element of $\mh$,
since its norm is $\cform{\sum}{k\geq 1}{}1=\infty$. The explanation
is that the $2$-point function changes during the deformation, and
so therefore does the inner product. Hence, if we Hilbert-complete,
the Hilbert space will change as well. 

\vspace{3mm}
For various reasons however we find this type of direct
approach difficult here. For one thing, we wish to consider
theories which really do not have Hilbert axiomatizations in the
proper sense, including
Minkowski signature theories, where the Hilbert approach is impossible
for physical reasons. Therefore, we prefer a ``vertex operator algebra''
approach where we discard the Hilbert completion and restrict ourselves
to examining tree level amplitudes. One such axiomatization of 
such theories was given in \cite{huang} under the term ``full field
algebra''. In the present paper, however, we prefer to work from
scratch, listing the properties we will use explicitly, and referring
to our objects as conformal field theories in the vertex operator
formulation. 

\vspace{3mm}
We will then consider untopologized vector spaces
\beg{einf10}{V=\bigoplus V_{(w_L,w_R)}.
}
Here $(w_L,w_R)$ are weights (we refer to $w_L$ resp. $w_R$
as the left resp. right component of the weight), so we assume $w_L-w_R\in \Z$ and
usually
\beg{einf10a}{w_L,w_R\geq 0,}
\beg{einf11}{V_{(0,0)}=\langle U_D\rangle.
}
The ``no ghost'' assumptions \rref{einf10a},
\rref{einf11} will sometimes be dropped. If there is a Hilbert space $\mh$,
then $V$ is interpreted as the ``subspace of states of finite weights''.
We assume that for $u\in V_{w_L,w_R}$, we have vertex operators of the form
\beg{einf12}{Y(u,z,\overline{z})=
\cform{\sum}{(v_L,v_R)}{}u_{-v_L-w_L,-v_R-w_R}z^{v_L}\overline{z}^{v_R}.
}
Here $u_{a,b}$ are operators which raise the left (resp.
right) component of weight by $a$ (resp. $b$). 
We additionally assume $v_L-v_R\in\Z$ and that for a given $w$, the weights
of operators which act on $w$ are discrete. Even more strongly,
we assume that 
\beg{einf13}{Y(u,z,\overline{z})=\cform{\sum}{i}{}Y_i(u,z)\tilde{Y}_i(u,\overline{z})
} 
where 
\beg{einf13a}{\begin{array}{l}Y_i(u,z)=\sum u_{i;-v_L-w_L}z^{v_L},\\
\tilde{Y}_i(u,\overline{z})=\sum \tilde{u}_{i;-v_R-w_R}\overline{z}^{v_R}
\end{array}
}
where all the operators $Y_i(u,z)$ commute with all $\tilde{Y}_j(v,\overline{z})$.
The main axiom \rref{einf12} must satisfy is ``commutativity'' and ``associativity''
analogous to the case of vertex operator algebras, i.e. there must exist
for fields $u,v,w\in V$ and $w^{\prime}\in V^\vee$ of finite
weight, a ``$4$-point function''
\beg{einfz}{w^{\prime}Z(u,v,z,\overline{z},t,\overline{t})w}
which is real-analytic and unbranched outside the loci of $z=0$, $t=0$
and $z=t$, and whose expansion in $t$ first and $z$ second (resp. $z$ first
and $t$ second, resp. $z-t$ first and $t$ second) is
$$w^{\prime}
Y(u,z,\overline{z})Y(v,t,\overline{t})w,$$ 
$$w^{\prime}Y(v,t,\overline{t})Y(u,z,\overline{z})w,$$
$$w^{\prime}Y(Y(u,z-t,\overline{z-t})v,t,\overline{t})w,$$ 
respectively. Here, for example, by an expansion in $t$ first and $z$
second we mean a series in the variable $z$ whose coefficients are series
in the variable $t$, and the other cases are analogous. 

\vspace{3mm}
We also assume that Virasoro algebras $\langle L_n\rangle$, $\langle \tilde{L}_n
\rangle$ with equal central charges $c_L=c_R$ act and that
\beg{einf14}{
\begin{array}{l}
Y(L_{-1}u,z,\overline{z})=\frac{\partial}{\partial z} Y(u,z,\overline{z}),\\
Y(\tilde{L}_{-1}u,z,\overline{z})=\frac{\partial}{\partial \overline{z}} 
Y(u,z,\overline{z})
\end{array}
}
and
\beg{einf15}{
\text{$V_{w_L,w_R}$ is the weight $(w_L,w_R)$ subspace of $(L_0,\tilde{L}_0)$.}
}

\vspace{3mm}
\noindent
{\bf Remark:}
Even the axioms outlined here are meant
for theories which are initial points of the proposed 
perturbative deformations, they are two restrictive for the theories
obtained as a result of the deformations themselves. To capture those
deformations, it is best to revert to Segal's approach, restricting
attention to genus $0$ worldsheets with a unique outbound boundary
component (tree level amplitudes). 
Operators will then be expanded both in the weight grading
and in the perturbative parameter (i.e. the coefficient
at each power of the deformation parameter will be an element
of the product-completed state space of the original
theory). To avoid discussion of topology,
we simply require that perturbative coefficients of all compositions of such
operators converge in the product topology with respect to the weight grading,
and the analytic topology in each graded summand.

\vspace{3mm}
In this section, we discuss infinitesimal perturbations, i.e. the
deformed theory is defined over $\C[\epsilon]/(\epsilon^2)$ where
$\epsilon$ is the deformation parameter. One case where such infinitesimal
deformations can be described explicitly is the following

\begin{theorem}
\label{l1}
Consider fields $u,v,w\in V$ where $u$ is 
primary of weight $(1,1)$. 
Next, assume that 
$$Z(u,v,z,\overline{z},t,\overline{t})=\cform{\bigoplus}{\alpha,\beta}{}
Z_{\alpha,\beta}(u,v,z,\overline{z},t,\overline{t})$$
where 
$$Z_{\alpha,\beta}(u,v,z,\overline{z},t,\overline{t})=
\cform{\bigoplus}{i}{}Z_{\alpha,\beta,i}(u,v,z,t)\tilde{Z}_{\alpha,\beta,i}
(u,v,\overline{z},\overline{t})$$
and for $w^{\prime}\in W^{\vee}$ of finite
weight, $w^{\prime}Z_{\alpha,\beta,i}(u,v,z,t)(z-t)^{\alpha}z^{\beta}$ 

\noindent
(resp. $w^{\prime}\tilde{Z}_{\alpha,\beta,i}
(u,v,\overline{z},\overline{t})\overline{z-t}^{\alpha}\overline{z}^{\beta}$) 
is a meromorphic (resp. antimeromorphic) function of $z$ on $\C P^{1}$, with poles
(if any) only at $0,t,\infty$.
Now write
\beg{einfl*}{Y_{u,\alpha,\beta}(v,t,\overline{t})=
(i/2)\int_{\Sigma}Z_{\alpha,\beta}(u,v,z,\overline{z},t,\overline{t})dzd\overline{z},
}
so 
$$Y_u(v,t,\overline{t})=Y(v,t,\overline{t}) +\epsilon \cform{\sum}{\alpha,\beta}{}
Y_{u,\alpha,\beta}(v,t,\overline{t})$$
is the infinitesimally deformed vertex operator where $\Sigma$ is the
degenerate worldsheet with unit disks cut out around $0,t,\infty$.
Assume now further that we can expand
\beg{einfl1}{Z_{\alpha,\beta,i}(u,v,z,t)=Y_{\alpha,\beta,i}(v,t)Y_{\alpha,\beta,i}(u,z)
\;
\text{when $z$ is near $0$},
}
\beg{einfl2}{Z_{\alpha,\beta,i}(u,v,z,t)=Y^{\prime}_{\alpha,\beta,i}(u,z)Y_{
\alpha,\beta,i}(v,t)
\;
\text{when $z$ is near $\infty$},
}
\beg{einfl3}{Z_{\alpha,\beta,i}(u,v,z,t)=Y_{\alpha,\beta,i}
(Y^{\prime\prime}_{\alpha,\beta,i}(u,z-t))v,
t)\;
\text{when $z$ is near $t$}.
}
Write
$$Y_{\alpha,\beta,i}(u,z)=\sum u_{\alpha,\beta,i,-n-\beta}z^{n+\beta-1},$$
$$Y_{\alpha,\beta,i}^{\prime}(u,z)=\sum u_{\alpha,\beta,i,-n-\alpha-\beta}^{\prime}
z^{n+\alpha+\beta-1},$$
$$Y_{\alpha,\beta,i}^{\prime\prime}(u,z)=\sum u_{\alpha,\beta,i,n-\alpha}^{\prime
\prime}z^{n+\alpha-1},$$
(Analogously with the $\tilde{}$'s.) 
Assume now
\beg{ell*}{u_{\alpha,\beta,i,0}w=0,\; u_{\alpha,\beta,i,0}^{\prime\prime}v=0,\; 
u_{\alpha,\beta,i,0}^{\prime}Y_{\alpha,\beta,i}(v,t)w=0}
and analogously for the $\tilde{}$'s (note that these conditions
are only nontrivial when $\beta=0$, resp. $\alpha=0$, resp. $\alpha=-\beta$).
Denote now by $\omega_{\alpha,\beta,i,0}$, 
$\omega_{\alpha,\beta,i,\infty}$,
$\omega_{\alpha,\beta,i,t}$ the indefinite integrals of 
\rref{einfl1}, \rref{einfl2}, \rref{einfl3}
in the variable $z$, obtained using the formula
$$\int z^kdz=\frac{z^{k+1}}{k+1}\; k\neq -1$$
(thus fixing the integration constant), and analogously with the $\tilde{}$'s.
Let then
\beg{einfl4}{
\begin{array}{l}
C_{\alpha,\beta,i}=\omega_{\alpha,\beta,i,\infty}-\omega_{\alpha,\beta,i,t},\\
D_{\alpha,\beta,i}=\omega_{\alpha,\beta,i,\infty}-\omega_{\alpha,\beta,i,0},\\
\tilde{C}_{\alpha,\beta,i}=\tilde{\omega}_{\alpha,\beta,i,\infty}-\tilde{\omega}_{
\alpha,\beta,i,t},\\
\tilde{D}_{\alpha,\beta,i}=\tilde{\omega}_{\alpha,\beta,i,\infty}-\tilde{\omega}_{
\alpha,\beta,i,0}
\end{array}
}
(see the comment in the proof on branching). Let
$$\phi_{\alpha,\beta,i}=
\pi \cform{\sum}{n}{}\frac{u_{\alpha,\beta,i,-n}\tilde{u}_{\alpha,
\beta,i,-n}}{n}$$
where 
$$Y_{\alpha,\beta,i}(u,z)=\sum u_{\alpha,\beta,i,-n}z^{n-1}$$
and similarly for the $\tilde{}$'s, the ${}^{\prime}$'s and the
${}^{\prime\prime}$'s. (The definition makes sense when
applied to fields on which the term with denominator $0$ 
vanishes.) Then
\beg{einfl5}{{
\begin{array}{l}\protect
Y_{\alpha,\beta,u}(v,t,\overline{t})w=\cform{\sum}{i}{}
\phi_{\alpha,\beta,i}^{\prime} Y(v,t,\overline{t})w \\
\protect
-Y(\phi_{\alpha,\beta,i}^{\prime\prime}v,t,\overline{t})w
-Y(v,t,\overline{t})\phi_{\alpha,\beta,i}w+\\
\protect
C_{\alpha,\beta,i}\tilde{C}_{\alpha,\beta,i}
(-1+e^{-2\pi i\alpha})+D_{\alpha,\beta,i}\tilde{D}_{\alpha,\beta,i}(1-e^{2\pi i
\beta}).
\end{array}
}}
Additionally, when $\alpha=0$, then $D_{\alpha,\beta,i}=\tilde{D}_{\alpha,\beta,i}=0$,
and when $\beta=0$ then $C_{\alpha,\beta,i}=\tilde{C}_{\alpha,\beta,i}=0,$
and
\beg{einfl6}{\protect
Y_{\alpha,\beta,u}(v,t,\overline{t})w=
\cform{\sum}{i}{}\phi_{\alpha,\beta,i}^{\prime} Y(v,t,\overline{t})w 
-Y(\phi_{\alpha,\beta,i}^{\prime\prime}v,t,\overline{t})w-Y(v,t,\overline{t})\phi_{\alpha,\beta,i}w.}
The equation \rref{einfl6} is also valid when $\alpha=-\beta$.

\end{theorem}

{\bf Remark 1:} Note that technically, the integral  \rref{einfl*} is
not defined on the nondegenerate worldsheet described. This can be
treated in the standard way, namely by considering an actual worldsheet
$\Sigma^{\prime}$ obtained by gluing on standard annuli on the boundary
components. It is easily checked that if we denote by $A^{u}_{q}$
the infinitesimal deformation of $A_q$ by $u$, then
$$A_{q}^{u}(w)=\phi A_q(w)- A_q(\phi w).$$
Therefore, the Theorem can be stated equivalently for the worldsheet 
$\Sigma^{\prime}$. The only change
needs to be made in formula \rref{einfl5}, where
$\phi^{\prime\prime}$ needs to be multiplied by $s^{-2n}$ and 
$\phi$ needs to be multiplied by $r^{-2n}$ where $r$ and $s$ are radii
of the corresponding boundary components. 
Because however this is equivalent, we can pretend
to work on the degenerate worldsheet $\Sigma$ directly, in
particular avoiding inconvenient scaling factors in the statement. 

\vspace{3mm}
\noindent
{\bf Remark 2:}
The validity of this Theorem is rather restricted by its assumptions.
Most significantly, its assumption states that the chiral $4$ point
function can be rendered meromorphic in one of the variables by
multiplying by a factor of the form $z^{\alpha}(z-t)^{\beta}$.
This is essentially equivalent to the fusion rules being ``abelian'',
i.e. $1$-dimensional for each pair of labels, and each pair of
labels has exactly one product. As we will see (and as is well known),
the $N=2$ minimal model is an example of a ``non-abelian'' theory.

Even for an abelian theory, the theorem only calculates the deformation
in the ``$0$ charge sector'' because of the assumption \rref{ell*}. Because
of this, even for a free field theory, we will need to discuss an extension
of the argument. Since in that case, however, stating precise
assumptions is even more complicated, we prefer to
treat the special case only, and to postpone the discussion to
Section \ref{sfree} below.

\Proof
Let us work on the scaled real worldsheet $\Sigma^{\prime}$. Let
$$\eta_{\alpha,\beta,i}=Z_{\alpha, \beta,i}(u,v,z,t)dz,$$
$$\tilde{\eta}_{\alpha,\beta,i}=\tilde{Z}_{\alpha,
\beta,i}(u,v,\overline{z},\overline{t})d\overline{z}.$$
Denote by $\partial_0$, $\partial_{\infty}$, $\partial_t$ the boundary
components of $\Sigma^{\prime}$ near $0$, $\infty$, $t$. Then
the form $\omega_{\alpha,\beta,i,\infty}\tilde{\eta}_{\alpha,\beta,i}$ 
is unbranched on a domain
obtained by making a cut $c$ connecting $\partial_0$ and $\partial_t$.
We have
\beg{einfl7}{\oint_{\partial_t}\omega_{\alpha,\beta,i,t}\tilde{\eta}=-
Y(\phi_{\alpha,\beta,i} v,t,\overline{t})
}
\beg{einfl8}{\oint_{\partial_0}\omega_{\alpha,\beta,i,0}\tilde{\eta}=-
Y(\phi_{\alpha,\beta,i} v,t,\overline{t})\phi_{\alpha,\beta,i}.
}
But we want to integrate $\omega_{\alpha,\beta,i}\tilde{\eta}_{\alpha,\beta,i}$
over th boundary $\partial K$:
\beg{einfl9}{\begin{array}{c}
\oint_{\partial K}\omega_{\alpha,\beta,i}\tilde{\eta}_{\alpha,\beta,i}=\\
\oint_{\partial_t}\omega_{\alpha,\beta,i}\tilde{\eta}_{\alpha,\beta,i}+
\oint_{\partial_0}\omega_{\alpha,\beta,i}\tilde{\eta}_{\alpha,\beta,i}+
\oint_{\partial_\infty}\omega_{\alpha,\beta,i}\tilde{\eta}_{\alpha,\beta,i}\\
+\int_{c^+}\omega_{\alpha,\beta,i}\tilde{\eta}_{\alpha,\beta,i}+
\int_{c^-}\omega_{\alpha,\beta,i}\tilde{\eta}_{\alpha,\beta,i}
\end{array}
}
where $c^+$, $c^-$ are the two parts of $\partial K$ along the cut $c$, oriented
from $\partial_t$ to $\partial_0$ and back respectively. Before going further,
let us look at two points $x^+\in c^+$, $x^-\in c^-$ which project to
the same point on $c$. We have
$$\begin{array}{l}
C(e^{-2\pi i\alpha}-1)\tilde{\eta}(x^-)=\\
C\tilde{\eta}(x^+)-C\tilde{\eta}(x^-)=(\omega_t+C)\tilde{\eta}(x^+)-
(\omega_t+C)\tilde{\eta}(x^-)=\\
\omega_{\infty}\tilde{\eta}(x^+)-\omega_{\infty}\tilde{\eta}(x^-)=(\omega_0+D)
\tilde{\eta}(x^+)-(\omega_o+D)\tilde(x^{-})=\\
D\tilde{\eta}(x^+)-D\tilde{\eta}(x^{-})=D(e^{2\pi i\beta}-1)\tilde{\eta}(x^-)
\end{array}
$$
(the subscripts $\alpha,\beta,i$ were omitted throughout to simplify
the notation). This implies the relation
\beg{einfl10}{C_{\alpha,\beta,i}(e^{-2\pi i\alpha}-1)=D_{\alpha,\beta,i}(
e^{2\pi i \beta}-1).
}
{\bf Comment:} This is valid when the constants $C_{\alpha,\beta,i}$,
$D_{\alpha,\beta,i}$ are both taken at the point $x^-$; note that
since the chiral forms are branched, we would have to adjust the statement
if we measured the constants elsewhere. This however will not be of
much interest to us as in the present paper we are most interested
in the case when the constants vanish.

\vspace{3mm}
In any case, note that \rref{einfl10} implies $C_{\alpha,\beta,i}=0$ when
$\beta=0\mod\Z$ and $\alpha\neq 0\mod\Z$, and $D_{\alpha,\beta,i}=0$ when $\alpha=0
\mod\Z$
and $\beta\neq 0\mod\Z$. There is an anlogous relation to \rref{einfl10}
between $\tilde{C}_{\alpha,\beta,i}$, $\tilde{D}_{\alpha,\beta,i}$.
Note that when $\alpha=0=\beta$, all the forms in sight are unbranched,
and \rref{einfl6} follows directly. To treat the case $\alpha=-\beta$,
proceed analogously, but replacing $\omega_{\alpha,\beta,i,\infty}$
by $\omega_{\alpha,\beta,i,0}$ or $\omega_{\alpha,\beta,i,t}$. Thus,
we have finished proving \rref{einfl6} under its hypotheses.

\vspace{3mm}
Returning to the general case, let us study the right hand side
of \rref{einfl9}. Subtracting the first two terms from \rref{einfl7},
\rref{einfl8}, we get
\beg{einfl11}{\oint_{\partial_t}C_{\alpha,\beta,i}\tilde{\eta}_{\alpha,
\beta,i}, \; 
\oint_{\partial_0}D_{\alpha,\beta,i}\tilde{\eta}_{\alpha,
\beta,i},
}
respectively. On the other hand, the sum of the last two terms, looking
at points $x^+,x^-$ for each $x\in c$, can be rewritten as
\beg{einfl12}{\int_{c^+} C_{\alpha,\beta,i}(-e^{-2\pi i\alpha}+1)
\tilde{\eta}_{\alpha,\beta,i}=\int_{c^-}D_{\alpha,\beta,i}(-e^{2\pi i \beta}+1)
\tilde{\eta}_{\alpha,\beta,i}.
}
Now recall \rref{einfl4}. Choosing $\tilde{\omega}_{\alpha,\beta,i,\infty}$
as the primitive function of $\tilde{\eta}_{\alpha,\beta,i}$, we see that for the
end point $x$ of $c^-$,
\beg{einfl13}{\begin{array}{l}
\tilde{\omega}_{\alpha,\beta,i,\infty}(x^+)-\tilde{\omega}_{\alpha,\beta,i,\infty}(x^-)
=\\
\tilde{\omega}_{\alpha,\beta,i,t}(x^+)-\tilde{\omega}_{\alpha,\beta,i,t}(x^-)=\\
(e^{-2\pi i\alpha}-1)\tilde{\omega}_{\alpha,\beta,i,t}(x^{-})=\\
(e^{-2\pi i\alpha}-1)\tilde{\omega}_{\alpha,\beta,i,\infty}(x^-)+
(e^{-2\pi i\alpha}-1)\tilde{C}_{\alpha,\beta,i}.
\end{array}
}
Similarly, for the beginning point $y$ of $c^-$,
\beg{einfl14}{\begin{array}{l}
-\tilde{\omega}_{\alpha,\beta,i,\infty}(y^+)+\tilde{\omega}_{\alpha,\beta,i,\infty}(y^-)
=\\
-\tilde{\omega}_{\alpha,\beta,i,0}(y^+)+\tilde{\omega}_{\alpha,\beta,i,0}(y^-)=\\
-(e^{2\pi i\beta}-1)\tilde{\omega}_{\alpha,\beta,i,0}(y^{-})=\\
-(e^{2\pi i\beta}-1)\tilde{\omega}_{\alpha,\beta,i,\infty}(y^-)-
(e^{2\pi i\beta}-1)\tilde{D}_{\alpha,\beta,i}.
\end{array}
}
Then \rref{einfl13}, \rref{einfl14} multiplied by $C_{\alpha,\beta,i}$
are the integrals \rref{einfl11}, while the integral \rref{einfl12} is
\beg{einfl15}{-D_{\alpha,\beta,i}(1-e^{2\pi i\beta})\tilde{\omega}_{\alpha,
\beta,i,0}(y^-)+C_{\alpha,\beta,i}(1-e^{-2\pi i\alpha})\tilde{\omega}_{
\alpha,\beta,i,0}(x^-).
}
Adding this, we get
$$C_{\alpha,\beta,i}\tilde{C}_{\alpha,\beta,i}
(-1+e^{-2\pi i\alpha})+D_{\alpha,\beta,i}\tilde{D}_{\alpha,\beta,i}(1-e^{2\pi i
\beta}),$$
as claimed.
\qed

\vspace{10mm}

\section{Exponentiation of infinitesimal deformations}

\label{sexp}

Let us now look at primary weight $(1,1)$ fields $u$. We would like
to investigate whether the infinitesimal deformation of vertex operators
(more precisely worldsheet vacua
or string amplitudes) along $u$ indeed continues to a finite deformation,
or at least to perturbative level, as discussed in the previous section.
Looking again at the equation \rref{einf4}, we see that we have in
principle a series of obstructions similar to those of Gerstenhaber
\cite{gerst}, namely if we denote by
\beg{eexp1}{L_n(m)=\cform{\sum}{i=0}{m}L_{n}^{i}\epsilon^{i},\;
L_{n}^{0}=L_n
}
a deformation of the operator $L_n$ in $Hom(V,V)[\epsilon]/\epsilon^m$,
we must have
\beg{eexp2}{L_n(m)u(m)=0\in V[\epsilon]/\epsilon^{m+1} \;\text{for $n>0$}
}
\beg{eexp3}{L_0(m)u(m)=u(m)\in V[\epsilon]/\epsilon^{m+1}.
}
This can be rewritten as
\beg{eexp4}{\begin{array}{l}
L_n u^m=-\cform{\sum}{i\geq 1}{}L_{n}^{i}u^{m-i}\\
(L_0-1)u^{m}=-\cform{\sum}{i\geq 1}{}L_{0}^{i}u^{m-i}.
\end{array}
}
(Analogously for the $\tilde{}$'s. In the following, we
will work on the obstruction for the chiral part, the
antichiral part is analogous.)
At first, these equations seem very overdetermined. Similarly
as in the case of Gerstenhaber's obstruction theory, however, of
course the obstructions are of cohomological nature. If we denote
by $\mathcal{A}$ the
Lie algebra $\langle L_0-1,L_1,L_2,...\rangle$, then the system
\beg{eexp5}{\begin{array}{l}
L_n(m)u(m-1)\\
(L_0(m)-1)u(m-1)
\end{array}}
is divisible by $\epsilon^m$ in $V[\epsilon]/\epsilon^{m+1}$,
and is obviously a coboundary, hence a cocycle with respect to 
$\langle L_0(m)-1,L_1(m),...\rangle$. Hence, dividing by $\epsilon^m$,
we get a $1$-cocycle of $\mathcal{A}$. Solving \rref{eexp4} means
expressing this $\mathcal{A}$-cocycle as a coboundary.

\vspace{3mm}
In the absence of ghosts (=elements of negative weights), 
there is another simplification we may take
advantage of. Suppose we have a $1$-cocycle $c=( x_0,x_1,...)$
of $\mathcal{A}$. (In our applications, we will be interested
in the case when the $x_{i}$'s are given by \rref{eexp4}.)
Then we have the equations
$$L^{\prime}_{k}x_j-L^{\prime}_{j}x_k=(k-j)x_{j+k},$$
where $L^{\prime}_{k}=L_k$ for $k>0$, $L^{\prime}_{0}=L_0-1$.
In particular,
$$L^{\prime}_{k}x_0-L^{\prime}_{0}x_k=kx_k,$$
or
\beg{eexp6}{L_kx_0=(L_0+k-1)x_k\;\text{for $k>0$}.
}
In the absence of ghosts, \rref{eexp6} means that for $k\geq 1$, $x_k$
is determined by $x_0$ with the exception of the weight $0$ summand
$(x_1)_0$ of $x_1$. Additionally, if we denote the weight $k$ summand
of $y$ in general by $y_k$, then 
\beg{eexp6a}{c=dy} 
means
\beg{eexp7}{(x_0)_k=(k-1)y,
}
\beg{eexp8}{(x_0)_1=0.
}
The rest of the equation \rref{eexp6a} then follows from \rref{eexp6},
with the exception of the weight $0$ summand of $x_1$. We must, then,
have
\beg{eexp9}{(x_1)_0\in Im L_1.
}
Conditions \rref{eexp8}, \rref{eexp9}, for
$$x_k=-\cform{\sum}{i\geq 1}{}L_{k}^{i}u^{m-i},$$
are the conditions for solving \rref{eexp4}, i.e. the actual obstruction.

\vspace{3mm}
For $m=1$, we get what we call the primary obstruction. We have
$$L_{k}^{1}=\tilde{L}_{-k}^{1}=\cform{\sum}{m,i}{} u_{i,m+k}\tilde{u}_{i,m},$$
so \rref{eexp8} becomes
\beg{eexp10}{\cform{\sum}{i}{}u_{i,0}\tilde{u}_{i,0}u=0.
}
The condition \rref{eexp9} becomes
\beg{eexp11}{\cform{\sum}{i}{}u_{i,1}\tilde{u}_{i,0}u\in Im L_1, 
\;\cform{\sum}{i}{}u_{i,0}\tilde{u}_{i,1}u\in Im \tilde{L}_1.
}

\vspace{3mm}
This investigation is also interesting in the supersymmetric context.
In the case of $N=1$ worldsheet supersymmetry, we have additional operators
$G^{i}_{r}$, and in the $N=2$ SUSY case, we have operators
$G^{+i}_{r}$, $G^{-i}_{r}$, $J^{i}_{n}$ (cf. \cite{greene,Nconf}),
defined as the $\epsilon^i$-coefficient of the deformation of $G_r$,
resp. $G^{+}_{r}$, $G^{-}_{r}$, $J_{n}$ analogously to equation \rref{eexp1}.

In the $N=1$-supersymmetric case, the critical deforming fields have
weight $(1/2,1/2)$ (as do $a$- and $c$-fields in the $N=2$ case), 
so in both cases the first equation \rref{eexp1} remains the
same as in the $N=0$ case, the second becomes
\beg{ecor39a}{(L_0-1/2)u^m=-\cform{\sum}{i\geq 1}{}L_{0}^{i}u^{m-i}.
}
Additionally, for $N=1$, we get
\beg{ecor39b}{G_r u^m=-\cform{\sum}{i\geq 1}{}G^{i}_{r}u^{m-i},\; r\geq 1/2
}
(similarly when $\tilde{\;}$'s are present).

In the $N=1$-supersymmetric case, we therefore deal with the Lie
algebra $\mathcal{A}$, which is the free $\C$-vector space on
$L_n$, $G_r$, $n\geq 0$, $r\geq 1/2$. For a cocycle which has
value $x_k$ on $L_k$ and $z_r$ on $G_r$, the equation \rref{eexp6}
becomes 
\beg{ecor41a}{L_k x_0=(L_0+k-1/2)x_k\;\text{for $k>0$},
}
so in the absence of ghosts, $x_k$ is always determined by $x_0$.
If
\beg{ecor42a}{
\text{the $1$-cocycle $(x_k,z_r)$ is the coboundary of $y$}
}
we additionally get
$$(x_0)_k=(k-1/2)y,$$
so
$$(x_0)_{1/2}=0.$$
On the other hand, on the $z$'s, we get
\beg{ecorA1}{G_r x_0=(L_0+r-1/2)z_r,\; r\geq 1/2,
}
so we see that in the absence of ghosts, all $z_r$'s are determined,
with the exception of 
$$(z_{1/2})_0.$$
Therefore our obstruction is
\beg{ecorA2}{(z_0)_{1/2}=0,\;(z_{1/2})_0\in Im(G_{1/2}).
}
For the primary obstruction, we have
\beg{ecorA3}{L^{1}_{k}=\tilde{L}^{1}_{-k}=
\cform{\sum}{m}{}(G_{-1/2}\tilde{G}_{-1/2}u)_{m+k,m}),
}
\beg{ecorA4}{
\begin{array}{l}G^{1}_{r}=2\cform{\sum}{m}{}(\tilde{G}_{-1/2}u)_{m+r,m},\\
\tilde{G}^{1}_{r}=2\cform{\sum}{m}{}(G_{-1/2}u)_{m,m+r},
\end{array}
}
so the obstruction becomes
\beg{ecorA5}{\begin{array}{l}
\cform{\sum}{m}{}(G_{-1/2}\tilde{G}_{-1/2}u)_{m,m}=0,\\
\cform{\sum}{m}{}(\tilde{G}_{-1/2}u)_{m+1/2,m}\in Im(G_{1/2}),\\
\cform{\sum}{m}{}(G_{-1/2}u)_{m,m+1/2}\in Im(\tilde{G}_{1/2}).
\end{array}
}
In the case of $N=2$ supersymmetry, there is an additional
complication, namely chirality. This means that in addition
to the conditions
\beg{ecorB1}{\begin{array}{l}
(L_0-1/2)u=0,\\
L_n G^{\pm}_{r}u=J_{n-1}=0 \;\text{for}\; n\geq 1, r\geq 1/2,
\end{array}
}
we require that $u$ be chiral primary, which means
\beg{ecorB2}{G^{+}_{-1/2}u=0.
}
(There is also the possibility of antichiral primary, which has
\beg{ecorB2'}{G^{-}_{-1/2}u=0
}
instead, and similarly at the $\tilde{\;}$'s.) Let us now write
down the obstruction equations for the chiral primary case.
We get the first equation \rref{eexp4}, \rref{ecor39a}, and
an analogue of \rref{ecor39b} with $G^{i}_{r}$ replaced
by $G^{+i}_{r}$ and $G^{-i}_{r}$. Additionally, we have the
equation
$$G^{+}_{-1/2}u^m=-\cform{\sum}{i\geq 1}{}G^{i}_{-1/2}u^{m-i}$$
and analogously for the $\tilde{\;}$'s.

In this situation, we consider the super-Lie algebra $\mathcal{A}_2$
which is the free $\C$-vector space on $L_n$, $J_n$, $n\geq 0$,
$G^{-}_{r}$, $r\geq 1/2$ and $G^{+}_{s}$, $s\geq -1/2$. One easily
verifies that this is a super-Lie algebra on which the central
extension vanishes canonically (\cite{greene}, Section 3.1).
Looking at a $1$-cocycle whose value is $x_k$,$z^{\pm}_{r}$, $t_k$
on $L_k$, $G^{\pm}_{r}$, $J_k$ respectively, we get the equation
\rref{ecor41a}, and additionally
\beg{ecorB3}{G^{\pm}_{r}x_0=(L_0+r-1/2)z^{\pm}_{r},
\;\text{$r\geq 1/2$ for $-$, $r\geq -1/2$ for $+$}
}
and
\beg{ecorB4}{J_n x_0=(n-1/2)t_n,\; n\geq 0.
}
We see that the cocycle is determined by $x_0$, with the exception of
$(z_{1/2}^{\pm})_0$, $(z_{-1/2}^{+})_1$. Therefore, we get the condition
\beg{ecorB5}{
\begin{array}{l}
(x_0)_{1/2}=0\\
(z^{\pm}_{1/2})_0\in Im(G^{\pm}_{1/2})\\
(z^{+}_{-1/2})_1=G^{+}_{-1/2}u\;\text{where $G^{+}_{1/2}u=0$}
\end{array}
}
and similarly for the $\tilde{\;}$'s.

In the case of deformation along a $cc$ field $u$, we have
\beg{ecorB6}{
L^{1}_{k}=\tilde{L}^{1}_{-k}=
\cform{\sum}{m}{} (G^{-}_{-1/2}\tilde{G}^{-}_{-1/2}u)_{m+k,m},
}
\beg{ecorB7}{\begin{array}{l}
G^{+,1}_{r}=\cform{\sum}{m}{}2(\tilde{G}^{-}_{-1/2}u)_{m+r+1/2,m}\\
\tilde{G}^{+,1}_{r}=\cform{\sum}{m}{}2(G^{-}_{-1/2}u)_{m,m+r+1/2}\\
G^{-,1}_{r}=\tilde{G}^{-,1}_{r}=0\\
J^{1}_{n}=0=\tilde{J}^{1}_{n},
\end{array}
}
so the obstructions are, in a sense, analogous to \rref{ecorA5} with $G_r$ replaced
by $G^{-}_{r}$.

\vspace{3mm}
\noindent
{\bf Remark:} The relevant computation in verifying that \rref{ecorB6},
\rref{ecorB7} (and the analogous cases before) form a cocycle uses
formulas of the following type (\cite{zhu}):
\beg{ecorC1}{Res_{z}(a(z)v(w)z^n)-Res_{z}(v(w)a(z)z^n)=
Res_{z-w}((a(z-w)v)(w)z^n).
}
For example, when $v$ is primary of weight $1$, $a=L_{-2}$,
the right hand side of \rref{ecorC1} is
$$
\begin{array}{l}
Res_{z-w}(L_0v(z-w)^{-2}n(z-w)w^{n-1}+L_{-1}v(z-w)^{-1}w^n)\\
=nv(w)w^{n-1}+L_{-1}v(w)w^n\\
=\sum nv_k w^{n-k-2}+\sum (-k-1)v_k w^{n-k-2}\\
=\sum (n-k)v_n w^{n-k-2}.
\end{array}
$$
The left hand side is $\sum [L_{n-1},v_{k-n+1}]w^{n-k-2}$, so 
we get
$$[L_{n-1}, v_{k-n+1}]=(n-k-1)v_k,$$
as needed. 

Other required identities follow in a similar way. Let us 
verify one interesting case when $a=G^{-1}_{-3/2}$,
$u$ chiral primary. Then the right hand side of \rref{ecorC1} is
$$Res_{z-w}(G^{-}_{-1/2}v(w)(z-w)^{-1}w^n)=(G^{-}_{-1/2}v)(w)=
\sum (G^{-}_{-1/2}v)w^{-n-1}.$$
This implies 
\beg{ek3q*}{[G^{-}_{r},u_s]=(G^{-}_{-1/2}u)_{r+s},}
as needed. 

\vspace{3mm}
We have now analyzed the primary obstructions for
exponentiation of infinitesimal CFT deformations. 
However, in order for a perturbative exponentiation to exist,
there are also higher obstructions which must vanish.
The basic principle for obtaining these obstructions
was formulated above. However, in pratice, it may often
happen that those obstructions will not converge. 
This may happen for two different basic reasons.
One possibility is that the deformation of the deforming
field itself does not converge. This is essentially a violation
of perturbativity, but may in some cases be resolved by
regularizing the CFT anomaly along the deformation parameter. We will discuss
this at the end of this section, and will give an example
in Section \ref{sfree} below. 

\vspace{3mm}
Even if all goes well with the parameter, however, there
may be another problem, namely the expressions for $L^{i}_{n}$
etc. may not converge due to the fact that our deformation
formulas concern vacua of actual worldsheets, while
$L^{i}_{n}$ etc. correspond to degenerate worldsheets. Similarly,
vertex operators may not converge in the deformed theories.
We will show here how to deal with this problem.

The main strategy is to rephrase the conditions from the above
part of this section in terms of ``finite annuli''. We start
with the $N=0$ (non-supersymmetric) case. Similarly as in \rref{eexp1},
we can expand
\beg{ecorf1}{U_{A_r}(m)=\cform{\sum}{h=0}{m}U_{A_{r}}^{h}\epsilon^h.
}
In the non-supersymmetric case, the basic fact we have is the following:

\begin{theorem}
\label{t1}
Assuming $u^k$ (considered as fields in the
original undeformed CFT) have weight $>(1,1)$ for $k<h$, $r\in (0,1)$
we have
\beg{ecorft1}{\begin{array}{l}
U^{h}_{A_r}=\\
\cform{\sum}{m_k}{}\cform{\int}{s_h=r}{1}s_{h}^{2m_h-1}\cform{\int}{s_{h-1}=r}{s_h}
s_{h-1}^{2m_{h-1}-1}...\\
\cform{\int}{s_{1}=r}{s_2}
s_{1}^{2m_{1}-1}u_{m_h,m_h}....u_{m_1,m_1}ds_1...ds_h U_{A_r}.
\end{array}
}
\beg{ecorft2}{u^h=\cform{\sum}{m_k}{}
\frac{1}{2^h(m_h+...+m_1)(m_{h-1}+...+m_1)...m_1}u_{m_h,m_h}...u_{m_1,m_1}u.
}
In particular, the obstruction is the vanishing of the sum (with the term
$m_h+...m_1$ omitted from the denominator) of the
terms in \rref{ecorft2} with $m_h+...m_1=0$.
\end{theorem}

\Proof
The identity \rref{ecorft1} is essentially by definition. The key point
is that in the higher deformed vacua, there are terms in the
integrand obtained by
inserting $u_k$, $k>1$ to boundaries of disjoint disks $D_i$ cut out of $A_r$.
Then there are corrective terms to be integrated on the worldsheets
obtained by cutting out those disks. But the point is that under our
weight assumption, all the disks $D_i$ can be shrunk to a single point,
at which point the term disappears, and we are left with integrals
of several copies of $u$ inserted at different points. If we are using
vertex operators to express the integral, the operators must additionally
be applied in time order (i.e. fields at points of lower
modulus are inserted first). There is an $h!$ permutation factor which cancels
with the Taylor denominator. This gives \rref{ecorft1}.

Now \rref{ecorft2} is proved by induction. For $h=1$, the calculation
is done above. Assuming the induction hypothesis, the term
of the integral where the $k-1$ innermost integrals have the
upper bound and the $k$'th innermost integral has the lower bound
is equal to
$$U^{h-k}_{A_r}u^k,\; h>k\geq 1.$$
The summand which has all upper bounds except in the last
integral is equal to
\beg{enpert+}{\frac{1-r^{2(m_1+...+m_h)}}{2^h
(m_h+...+m_1)(m_{h-1}+...+m_1)...m_1}u_{m_h,m_h}...u_{m_1,m_1}ur^2,
}
which is supposed to be equal to
$$-U_{A_r}u^h +r^2u^h.$$
This gives the desired solution.
\qed

\vspace{3mm}
\noindent
{\bf Remark:}
The formula \rref{enpert+} of course does not apply to the case
$m_1+...+m_h=0$. In that case, the correct formula is
\beg{enpert++}{\frac{-\ln(r)}{
(m_{h-1}+...+m_1)...m_1}u_{m_h,m_h}...u_{m_1,m_1}ur^2.
}
So the question becomes whether there could exist a field $u^h$ such that
$U_{A_r}u^h-r^2 u^h$ is equal to the quantity \rref{enpert++}.
One sees immediately that such field does not exist in the product-completed
space of the original theory. What this approach does not settle however
is whether it may be possible to add such non-perturbative fields
to the theory and preserve CFT axioms, which could facilitate
existence of deformations in some generalized
sense, despite the algebraic obstruction. It would have
to be, however, a field of generalized weight in the sense of 
\cite{zhang,zhang1,huang,huang1}. 

In effect, written in infinitesimal terms, the relation 
\rref{enpert++} becomes
$$L_0u^h-u^h=-\frac{1}{(m_{h-1}+...+m_1)...m_1u}_{m_h,m_h}...u_{m_1,m_1}u.
$$
The right hand side $w$ is a field of holomorphic weight $1$, so
we see that we have a matrix relation
$$L_0\left(\begin{array}{l}u^h\\u\end{array}
\right)=\left(\begin{array}{rr}1 &w\\0&1\end{array}
\right)\left(\begin{array}{l}u^h\\u\end{array}
\right)
$$
This is an example of what one means by a field of generalized weight.
One should note, however, that fields of generalized weight
are excluded in unitary conformal field theories. By Wick
rotation, the unitary axiom of a conformal field theory
becomes the axiom of reflection positivity \cite{scft}: 
the operator $U_\Sigma$ associated with a worldsheet $\Sigma$ is defined
up to a $1$-dimensional complex line $L_\Sigma$ (which is often
more strongly assumed to have a positive real structure). 
If we denote by $\overline{\Sigma}$ the complex-conjugate worldsheet
(note that this reverses orientation of boundary components),
then reflection positivity requires that we have an isomorphism
$L_{\overline{Sigma}}\cong L_{\Sigma}^{*} 
$ (the dual line),
and using this isomorphism, an identification $U_{\overline{\Sigma}}=
U_{\Sigma}^{*}$ (here the asterisk denotes the adjoint operator). 
Specializing to annuli $A_r$, $||r||\leq 1$, we see that the annulus
for $r$ real is self-conjugate, so the corresponding operators
are self-adjoint, and hence diagonalizable. On the other hand,
for $||r||=1$, we obtain unitary operators, and unitary representations
of $S^1$ on Hilbert space split into eigenspaces of integral
weights. The central extension given by $L$ is then trivial and
hence the operators corresponding to all $A_r$ commute, and hence
are simultaneously diagonalizable, thus excluding the possibility
of generalized weight. 

The possibility, of course, remains that the correlation function
of the deformed theory can be modified by a non-perturbative correction.
Let us note that if left uncorrected, the term \rref{enpert++}
can be interpreted infinitesimally as
\beg{eiint1}{L_0u(\epsilon)-u(\epsilon)= C\epsilon^m v\mod 
\epsilon^{m+1},
}
where $v$ is another field of weight $1$. Note that in case 
that $u=v$, \rref{eiint1} can be interpreted as saying that $u$
changes weight at order $m$ of the perturbation
parameter. In the general case, we obtain a matrix involving
all the (holomorphic) weight $1$ fields in the unperturbed 
theory. Excluding fields of generalized weight in the unperturbed
theory (which would translate to fields of generalized weight
in the perturbed theory), the matrix must have other eigenvalues
than $1$, thus showing that some critical fields will change weight.

\vspace{3mm}
In the $N=1$-supersymmetric case, an analogous statement holds, 
except the assumption is that the weight of $u^k$ is greater than
$(1/2,1/2)$ for $k<h$, and the integral \rref{ecorft1} must
be replaced by
\beg{ecorft1a}{\begin{array}{l}
U^{h}_{A_r}=\\
\cform{\sum}{m_k}{}\cform{\int}{s_h=r}{1}s_{h}^{m_h-1}\cform{\int}{s_{h-1}=r}{s_h}
s_{h-1}^{m_{h-1}-1}...\\
\cform{\int}{s_{1}=r}{s_2}
s_{1}^{m_{1}-1}(G_{-1/2}\tilde{G}_{-1/2}u)_{m_h,m_h}...
(G_{-1/2}\tilde{G}_{-1/2}u)_{m_1,m_1}
ds_1...ds_h U_{A_r},
\end{array}
}
and accordingly 
\beg{ecorft2a}{
\begin{array}{l}
u^h=\cform{\sum}{m_k}{}
\frac{1}{2^h(m_h+...+m_1)(m_{h-1}+...+m_1)...m_1}
\\(G_{-1/2}\tilde{G}_{-1/2}u)_{m_h,m_h}
...(G_{-1/2}\tilde{G}_{-1/2}u)_{m_1,m_1}u,
\end{array}
}
so the obstruction again states that the term with $m_h+...+m_1=0$ must
vanish. In the $N=2$ case, when $u$ is a $cc$ field, we simply replace
$G$ by $G^-$ in \rref{ecorft1a}, \rref{ecorft2a}.

But in the supersymmetric case, to preserve supersymmetry along the
deformation, we must also investigate the ``finite" analogs of the
obstructions associated with $G_{1/2}$ in the $N=1$ case, and
$G^{\pm}_{1/2}$, $G^{+}_{-1/2}$ in the $N=2$ $c$ case (and similarly for
the $a$ case, and the $\tilde{\;}$'s). In fact, to tell the whole
story, we should seriously investigate integration of the deforming
fields over super-Riemann surfaces (=super-worldsheets). This can
be done; one approach is to treat the case of the superdisk
first, using Stokes theorem twice with the differentials $\partial$,
$\overline{\partial}$ replaced by $D$, $\overline{D}$ respectively
in the $N=1$ case (and the same at one chirality for the $N=2$ case).
A general super-Riemann surface is then partitioned into superdisks.

For the purpose of obstruction theory, the following special case is
sufficient. We treat the $N=2$ case, since it is of main interest
for us. Let us consider the case of $cc$ fields
(the other cases are analogous). First we note (see \rref{ecorB7}) that $G^-$ is unaffected
by deformation via a $cc$ field, so the obstructions derived
from $G^{-}_{-1/2}$ and $G^{-}_{1/2}$ are trivial (and similarly
at the $\tilde{\;}$'s).

To understand the obstruction associated with $G^{+}_{1/2}$, we will 
study ``finite" (as opposed to infinitesimal) annuli obtained
by exponentiating $G^{+}_{1/2}$. Now the element $G^{+}_{1/2}$
is odd. Thinking of the super-semigroup of superannuli as a supermanifold,
then it makes no sense to speak of ``odd points" of the supermanifold.
It makes sense, however, to speak of a family of edd elements parametrized
by an odd parameter $s$: this is simply the same thing as a map
from the $(0|1)$-dimensional superaffine line into the supermanifold. 
In this sense, we can speak of the ``finite" odd annulus
\beg{eodd1}{\exp(s G^{+}_{1/2}).
}
Now we wish to study the deformations of teh operator associated
with \rref{eodd1} along a $cc$ field $u$ as a perturbative
expansion in $\epsilon$.

Thinking of $G^{+}_{1/2}$ as an $N=2$-supervector field, we have
\beg{eodd2}{G^{+}_{1/2}=(z+\theta^+\theta^-)\frac{\partial}{\partial\theta^+}-
z\theta^{-}\frac{\partial}{\partial z}.
}
We see that \rref{eodd2} deforms infinitesimally only the variables
$\theta^+$ and $z$, not $\theta^-$. Thus, more specifically,
\rref{eodd1} results in the transformation
\beg{eodd3}{
\begin{array}{l}
z\mapsto \exp(s\theta^-)z\\
\theta^-\mapsto\theta^-.
\end{array}
}
This gives rise to the formula, valid when $u^k$ have weight $>(1/2,1/2)$
for $1\leq k<h$, 
\beg{eodd4}{
\begin{array}{l}
U^{h}_{\exp(sG^{+}_{1/2})}=\\
\cform{\sum}{m_k}{}\cform{\int}{t_h=\exp(s\theta^-)}{1}t_{h}^{m_h-1}
\cform{\int}{t_{h-1}=\exp(s\theta^-)}{t_h}
t_{h-1}^{m_{h-1}-1}...\\
\cform{\int}{t_{1}=\exp(s\theta^-)}{t_2}
s_{1}^{m_{1}-1}v_{m_h,m_h}....v_{m_1,m_1}dt_1...dt_h U_{\exp(s G^{+}_{1/2})},
\end{array}
}
where $v_{m_k,m_k}$ is equal to 
\beg{eoddD1}{(\tilde{G}^{-}_{-1/2}u)_{m_{k+1/2},m_k}}
in summands of \rref{eodd4}
where the factor resulting from
integrating the $t_k$ variable has a $\theta^-$ factor, and
\beg{eoddD2}{(G^{-}_{-1/2}\tilde{G}^{-}_{-1/2}u)_{m_k,m_k}}
in other summands. (We see that each summand can be considered 
as a product of factors resulting from integrating the individual
variables $t_k$; in at most one factor, \rref{eoddD1} can occur,
otherwise the product vanishes.)

Realizing that $exp(ms\theta^-)=1+ms\theta^-$, this gives
that the obstruction (under the weight assumption for $u^k$)
is that the summand for $m_1+...+m_h=0$ (with the denominator
$m_1+...+m_h$ omitted) in the following
expression vanish:
\beg{eodd*}{
\begin{array}{l}
\cform{\sum}{m_k}{}\cform{\sum}{k=1}{h}\frac{1}{m_1+...+m_h}...
\frac{1}{m_1}\\
(G^{-}_{-1/2}\tilde{G}^{-}_{-1/2}u)_{m_h,m_h}...
m_k
(\tilde{G}^{-}_{-1/2}u)_{m_{k+1/2},m_k}...
(G^{-}_{-1/2}\tilde{G}^{-}_{-1/2}u)_{m_1,m_1}u.
\end{array}
}

\vspace{3mm}
To investigate the higher obstructions further, we need the language
of correlation functions. Specifically, the CFT's whose deformations
we will consider are ``RCFT's". The simplest way of building an RCFT
is from ``chiral sectors" $\mathcal{H}_{\lambda}$ where $\lambda$
runs through a set of labels, by the recipe
$$\mathcal{H}=\cform{\bigoplus}{\lambda}{}\mh_{\lambda}\otimes\mh_{\lambda^*}$$
where $\lambda^*$ denotes the contragredient label (cf. \cite{kondo}). 
(In the case of the Gepner model, we will need a slightly more general
scenario, but our methods still apply to that case analogously.)
Further, we will have a symmetric bilinear form
$$B:\mh_{\lambda}\otimes \mh_{\lambda^*}\r \C$$
with respect to which the adjoint to $Y(v,z)$ is
$$(-z^{-2})^n Y(e^{zL_1}v,1/z)$$
where $v$ is of weight $n$. There is also a real
structure
$$\mh_{\lambda}\cong\overline{\mh}_{\lambda^*},$$
thus specifying a real structure on $\mh$, $\overline{u\otimes v}=\overline{u}
\otimes \overline{v}$, and inner product
$$\langle u_1\otimes v_1,u_2\otimes v_2\rangle=B(u_1,\overline{u_2})B(v_1,
\overline{v_2}).$$
We also have an inner product
$$\mh_{\lambda}\otimes_{\R} \mh_{\lambda^*}\r \C$$
given by
$$\langle u,v\rangle=B(u,\overline{v}).$$
Then we have the $\P^1$-chiral correlation function
\beg{ecorel1}{\langle u(z_{\infty})^*|v_m(z_m)v_{m-1}(z_{m-1})...v_1(z_1)v_0(z_0)
\rangle
}
which can be defined by taking the vacuum operator associated with the
degenerate worldsheet $\Sigma$
obtained by ``cutting out'' unit disks with centers
$z_0,...,z_m$ from the unit disk with center $z_\infty$, applying this
operator to $v_0\otimes...\otimes v_m$, and taking inner product with
$u$. Thus, the correlation function \rref{ecorel1} is in fact the same
thing as applying the field on either side of \rref{ecorel1} to
the identity, and taking the inner product.

This object \rref{ecorel1} is however not simply a function of 
$z_0,..., z_\infty$. Instead, there is a finite-dimensional vector
space $M_\Sigma$ depending holomorphically on $\Sigma$ (called
the modular functor) such that \rref{ecorel1} is a linear function
$$M_{\Sigma}\r \C.$$
However, now one assumes that $M$ is a ``unitary modular functor"
in the sense of Segal \cite{scft}. This means that $M_\Sigma$
has the structure of a positive-definite inner product space
for not just the $\Sigma$ as above, but an arbitrary worldsheet.
The inner product is not valued in $\C$, but in 
$$||det(\Sigma)||^{2c}$$
where $c$ is the central charge. Since the determinant of $\Sigma$ as above
is the same as $det(\P^1)$ (hence in particular constant), we can
make the inner product $\C$-valued in our case. 

\vspace{3mm}
If the deforming field is of the form 
\beg{euotimes}{u\otimes\tilde{u},}
the ``higher $L_0$ obstruction" (under the weight assumptions given
above) can be further written as
\beg{ecorel+}{\begin{array}{l}
\cform{\int}{0\leq ||z_1||\leq||z_m||\leq 1}{}
\langle v(0)^*|u(z_m)...u(z_1)u(0)\rangle\\
\langle \tilde{v}^*|\tilde{u}(\overline{z_m})..\tilde{u}(\overline{z_1})\tilde{u}(0)\rangle
dz_1 d\overline{z}_1....dz_md\overline{z}_m\\
\text{for $w(v)\leq 1$}
\end{array}
}
($w$ is weight) in the $N=0$ case and
\beg{ecorel++}{\begin{array}{l}
\cform{\int}{0\leq ||z_1||\leq||z_m||\leq 1}{}
\langle v(0)^*|(G^{-}_{-1/2}u)(z_m)...(G^{-}_{-1/2}u)(z_1)u(0)\rangle\\
\langle \tilde{v}^*|(\tilde{G}^{-}_{-1/2}\tilde{u})(\overline{z_m})...
(\tilde{G}^{-}_{-1/2}\tilde{u})(\overline{z_1})\tilde{u}(0)\rangle
dz_1 d\overline{z}_1....dz_md\overline{z}_m\\
\text{for $w(v)\leq 1/2$}
\end{array}
}
in the $N=2$ $cc$ case. The $G^{+}_{1/2}$-obstruction in the $N=2$
case can be written as
\beg{ecorel+++}{\begin{array}{l}
\cform{\int}{0\leq ||z_1||\leq||z_m||\leq 1}{}
\cform{\sum}{k=1}{m}
\langle v(0)^*|(G^{-}_{-1/2}u)(z_m)...u(z_k)...(G^{-}_{-1/2}u)(z_1)u(0)\rangle\\
\langle \tilde{v}^*|(\tilde{G}^{-}_{-1/2}\tilde{u})(\overline{z_m})...
(\tilde{G}^{-}_{-1/2}\tilde{u})(\overline{z_1})\tilde{u}(0)\rangle\\
dz_1 d\overline{z}_1....dz_md\overline{z}_m\;\text{for $w(v)\leq 0$,
$w(\tilde{v})\leq 1/2$}
\end{array}
}
and similarly for the $\tilde{\;}$.
We see that these obstructions vanish when we have
\beg{ecoreli}{\langle v(z_\infty)^*|u(z_m)....u(z_0)\rangle=0,\;
\text{for $w(v)\leq 1$}
}
in the $N=0$ case (and similarly for the $\tilde{\;}$'s), and
\beg{ecorelii}{\langle v(z_\infty)^*|G^{-}_{-1/2}u(z_m)....G^{-}_{-1/2}u(z_1)
u(z_0)\rangle=0,\;
\text{for $w(v)\leq 1/2$,}
}
and similarly for the $\tilde{\:}$'s. Observe further that when
$$\tilde{u}=\overline{u},$$
the condition for the $\tilde{\:}$'s is equivalent to the condition
for $u$, and further \rref{ecoreli}, \rref{ecorelii} are
also necessary in this case, as in \rref{ecorel+}, \rref{ecorel++}
we may also choose $\tilde{v}=\overline{v}$, which makes the
integrand non-negative (and only $0$ if it is $0$ at each chirality).
In the $N=2$ case, it turns the condition \rref{ecorelii}
simplifies further:

\begin{theorem}
\label{t2}
Let $u$ be a chiral primary field of weight $1/2$. Then the necessary and
sufficient condition \rref{ecorelii} for existence of perturbative
CFT deformations along the field $u\otimes \overline{u}$ is equivalent
to the same vanishing condition applied to only chiral primary fields
$v$ of weight $1/2$.
\end{theorem}

\Proof

In order
for the fields \rref{ecorelii} to correlate, they would have to have
the same $J$-charge $Q_J$. We have
$$Q_J u=1,\; Q_J (G^{-}_{-1/2}u)=0.$$
As $Q_J$ of the right hand side of \rref{ecorelii} is $1$. Thus, for 
the function \rref{ecorelii} to be possibly non-zero, we must have
\beg{ecorelti}{Q_J v=1.
}
But then we have
$$w(v)\geq \frac{1}{2}Q_J v=\frac{1}{2}$$
with equality arising if and only if 
\beg{ecoreltii}{\text{$v$ is chiral primary of weight $1/2$.}}
\qed

\vspace{3mm}
\noindent
{\bf Remark 1:}
We see therefore that in the $N=2$ SUSY case, there is in fact no
need to assume that the weight of $U^k$ is $>(1/2,1/2)$ for 
$k<h$. If the obstruction vanishes for $k<h$, then we have
\beg{epert1}{u^k=\frac{1}{k!}\cform{\int}{D}{}
(G^{-}_{-1/2}\tilde{G}^{-}_{-1/2}u)(z_k)...
(G^{-}_{-1/2}\tilde{G}^{-}_{-1/2}u)(z_1)udz_1...dz_kd\overline{z}_1...
d\overline{z}_k
}
where the integrand is understood as a $(k+1)$-point function
(and not its power series expansion in any particular range),
over the unit disk. 

Additionally, for any worldsheet $\Sigma$,
\beg{epert2}{U^{h}_{\Sigma}=\frac{1}{h!}\cform{\int}{\Sigma}{}
(G^{-}_{-1/2}\tilde{G}^{-}_{-1/2}u)(z_h)...
(G^{-}_{-1/2}\tilde{G}^{-}_{-1/2}u)(z_1)dz_1...dz_hd\overline{z}_1...
d\overline{z}_h
}
(it is to be understood that in both \rref{epert1}, \rref{epert2}, the fields
are inserted into holomorphic images of disks where the origin maps to the
point of insertion with derivative of modulus 1 with respect to the measure
of integration).

When the obstruction occurs at step $k$, the integral \rref{epert1} has
a divergence of logarithmic type. In the $N=0$ case, there is a third
possibility, namely that the obstruction vanishes, but the field $u_h$
in Theorem \ref{t1} has summands of weight $<(1,1)$ ($<(1/2,1/2)$ for
$N=1$). In this case, the integral \rref{epert1} will have a divergence
of power type, and the intgral of terms of weight $<(1,1)$ (resp. $<(1/2,1/2)$)
has to be taken in range from $\infty$ to $1$ rather than from $0$ to $1$
to get a convergent integral. The formula \rref{epert2} is not correct
in that case. 

\vspace{3mm}
\noindent
{\bf Remark 2:} In \cite{dixon}, a different correlation function is
considered as an measure of marginality of $u$ to higher perturbative
order. The situation there is actually more general, allowing
combinations of both chiral and antichiral primaries. In the 
present setting of chiral primaries only, the correlation function
considered in \cite{dixon} amounts to
\beg{edixon}{\langle 1|(G^{-}_{-1/2}u)(z_n)...(G^{-}_{-1/2}u)(z_1)\rangle.
}
It is easy to see using the standard contour deformation argument to
show that \rref{edixon} indeed vanishes, which is also observed
in \cite{distler}. In \cite{dixon}, this type of vanishing
is taken as evidence that the $N=2$ CFT deformations exist. 
It appears, however, that even though the vanishing
of \rref{edixon} follows from the vanishing of \rref{ecorelii},
the opposite implication does not hold. (In fact, we will see
examples in Section \ref{sexam} below.) 
The explanation seems to be that \cite{dixon} writes
down an integral expressing the change of central charge
when deforming by a combination of cc fields and ac fields,
and proves its vanishing. While this is correct formally, 
we see from Remark 1 above that in fact a singularity
can occur in the integral when our obstruction is non-zero:
the integral can marginally diverge for $k$ points
while it is convergent for $<k$ points.

\vspace{3mm}
It would be nice if the obstruction theory a la Gerstenhaber we
described here settled in general the question of deformations of
conformal field theory, at least in the vertex operator formulation.
It is, however, not that simple. The trouble is that we are not in a 
purely algebraic situation. Rather, compositions of operators which
are infinite series may not converge, and even if they do, the convergence
cannot be understood in the sense of being eventually constant, but
in the sense of analysis, i.e. convergence of sequences of real numbers.

\vspace{3mm}
Specifically, in our situation, there is the possibility of
divergence of the terms on the right hand side of \rref{eexp4}.
Above we dealt with one problem, that in general, we do not expect infinitesimal
deformations to converge on the degenerate worldsheets of vertex operators,
so we may have to replace \rref{eexp4} by equations involving finite
annuli instead. However, that is not the only problem. We may encounter
{\em regularization along the flow parameter}. This stems from the
fact that the equations \rref{eexp2}, \rref{eexp3} only determine
$u(\epsilon)$ up to scalar multiple, where the scalar may be of the form
\beg{eexp20}{1+\cform{\sum}{i\geq 1}{}K_i\epsilon^i=f(\epsilon).
}
But the point is (as we shall see in an example in the next section)
that we may only be able to get a well defined value of
\beg{eexp21}{f^{-1}(\epsilon)u(\epsilon)=v(\epsilon)
}
when the constants $K_i$ are infinite. The obstruction then is
\beg{eexp22}{\begin{array}{l}
L_n(m)f(m)v(m)=0\in V[\epsilon]/\epsilon^{m+1}\;\text{for $n>0$}\\
(1-L_0(m))f(m)v(m)=0\in V[\epsilon]/\epsilon^{m+1}.
\end{array}
}
At first, it may seem that it is difficult to make this
rigorous mathematically with the infinite constants present.
However, we may use the followng trick. Suppose we want to
solve
\beg{eexp23}{\begin{array}{c}
c_1a_{11}+...+c_n a_{1n}=b_1\\
...\\
c_1a_{m1}+...+c_n a_{mn}=b_n
\end{array}
}
in a, say, finite-dimensional vector space $V$. Then we make rewrite \rref{eexp23}
as
\beg{eexp24}{(b_1,...,b_n)=0\in(\cform{\sum}{m}{} V)/
\langle(a_{11},...,a_{m1}),...,(a_{1n},...,a_{mn})\rangle.
}
This of course doesn't give anything new in the algebraic situation,
i.e when the $a_{ij}$'s are simply elements of the vector
space $V$. When, however the vectors 
$$(a_{11},...,a_{m1}),...,(a_{1n},...,a_{mn})$$
are (possibly divergent) infinite sums
$$(a_{1j},...,a_{mj})=\cform{\sum}{k}{}(a_{1jk},...,a_{mjk}),$$
then the right hand side of \rref{eexp24} can be interpreted as
$$(\cform{\sum}{m}{}V)/
\langle(a_{11k},...,a_{m1k}),...,(a_{1nk},...,a_{mnk})\rangle.$$
In that sense, \rref{eexp24} always makes sense, while \rref{eexp23}
may not when interpreted directly. We interpret \rref{eexp22} in this
way. 

\vspace{3mm}
Let us now turn to the question of sufficient conditions for
exponentiation of infinitesimal deformations. Suppose there exists
a subspace $W\subset V$ closed under vertex operators which 
contains $u$ and such that for all elements $v\in W$, we have
that
$$\cform{\sum}{i}{}Y_i(u,z)\tilde{Y}_i(u,\overline{z})v$$
involve only $z^n\overline{z}^m$ with $m,n\in\Z$, $m,n\neq -1$.
Then, by Theorem \ref{l1}, 
$$1-\phi\epsilon:W\r\hat{W}[\epsilon]/\epsilon^2$$
is an infinitesimal isomorphism between $W$ and the infinitesimally
$u$-deformed $W$. It follows, in the non-regularized case, that
then
\beg{eexp25}{\exp(-\phi\epsilon) u
}
is a globally deformed primary field of weight $(1,1)$, and
\beg{eexp26}{\exp(-\phi\epsilon):W
\r \hat{W}[[\epsilon]]
}
is an isomorphism between $W$ and the exponentiated deformation of
$W$. However, since we now know the primary fields along the deformation,
vacua can be recovered from the equation \rref{einf4} of the last
section. 

\vspace{3mm}
Such nonregularized exponentiation occurs in the case of the
{\em coset construction}. Setting
$$W=\langle v|\parbox{2.5in}{$Y_i(u,z)\tilde{Y}_i(u,\overline{z})v$
involve only $z^n\overline{z}^m$ with $m,n\geq 0$, $m,n\in\Z$}\rangle.$$
Then $W$ is called the coset of $V$ by $u$. Then $W$ is closed under
vertex operators, and if $u\in W$, the formulas \rref{eexp25},
\rref{eexp26} apply without regularization.

\vspace{3mm}
The case with regularization occurs when there exists some
constant
$$K(\epsilon)=1+\cform{\sum}{n\geq 1}{}K_n\epsilon^n$$
where $K_n$ are possibly constants such that
\beg{eexp27}{K(\epsilon)\exp(-\phi\epsilon)u
}
is finite in the sense described above (see \rref{eexp24}).
We will see an example of this in the next section.

\vspace{3mm}
All these constructions are easily adapted to supersymmetry.
The formulas \rref{eexp25}, \rref{eexp26} hold without
change, but the deformation is with respect to
$G_{-1/2}\tilde{G}_{-1/2}u$ resp. $G_{-1/2}^{-}\tilde{G}_{-1/2}^{-}u$,
$G_{-1/2}^{+}\tilde{G}_{-1/2}^{-}u$ depending on the situation
applicable.

\vspace{10mm}

\section{The deformations of free field theories}

\label{sfree}

\vspace{5mm}

As our first application, let us consider the $1$-dimensional bosonic
free field conformal field theory, where the deformation field
is
\beg{efree1}{u=x_{-1}\tilde{x}_{-1}.
}
In this case, the infinitesimal isomorphism of Theorem \ref{l1} satisfies
\beg{efree2}{ \phi=\pi\cform{\sum}{n\in\Z}{}\frac{x_{-n}\tilde{x}_{-n}}{n}
}
and the sufficient condition of exponentiability from the last section
is met when we take $W$ the subspace consisting of states of momentum $0$.
Then $W$ is closed under vertex operators, $u\in W$ and the $n=0$ term
of \rref{efree2} drops out in this case. However, this is an example
where regularization is needed. It can be realized as follows:
Write
$$\phi=\cform{\sum}{n>0}{}\phi_n$$
where
$$\phi_n=\pi(\frac{x_{-n}\tilde{x}_{-n}}{n}-\frac{x_{n}\tilde{x}_{n}}{n}).$$
We have
\beg{efree3}{\exp\phi=\cform{\prod}{n>0}{}\exp\phi_n.
}
To calculate $\exp\phi_n$ explicitly, we observe that
$$[\frac{x_{-n}\tilde{x}_{-n}}{n},\frac{x_{n}\tilde{x}_{n}}{n}]=
-\frac{x_{-n}x_{n}}{n}-\frac{\tilde{x}_{-n}\tilde{x}_{n}}{n}-1,$$
and setting
\beg{efree4}{\begin{array}{l}
e=\frac{x_{-n}\tilde{x}_{-n}}{n},\\
f=\frac{x_{n}\tilde{x}_{n}}{n},\\
h=-\frac{x_{-n}x_{n}}{n}-\frac{\tilde{x}_{-n}\tilde{x}_{n}}{n}-1,
\end{array}
}
we obtain the $sl_2$ Lie algebra
\beg{efree5}{\begin{array}{l}
[e,f]=h,\\
{}[e,h]=2e,\\
{}[f,h]=-2f.
\end{array}
}
Note that conventions regarding the normalization of $e,f,h$ vary, but
the relations \rref{efree5} are satisfied for example for
\beg{efree6}{e=\left(\begin{array}{rr}0&1\\0&0\end{array}\right),\;
f=\left(\begin{array}{rr}0&0\\-1&0\end{array}\right),\;
h=\left(\begin{array}{rr}-1&0\\0&1\end{array}\right).
}
In $SL_2$, we compute
\beg{efree7}{\begin{array}{l}
\exp\pi\epsilon(-f+e)=\exp\pi\epsilon
\left(\begin{array}{rr}0&1\\1&0\end{array}\right)\\
=\left(\begin{array}{rr}\cosh\pi\epsilon&\sinh\pi\epsilon
\\ \sinh\pi\epsilon&\cosh\pi\epsilon\end{array}\right)\\
\left(\begin{array}{cc}1&\tanh\pi\epsilon\\0&1\end{array}\right)
\left(\begin{array}{cc}\frac{1}{\cosh\pi\epsilon}&0\\0&\cosh\pi\epsilon
\end{array}\right)
\left(\begin{array}{cc}1&0\\\tanh\pi\epsilon&1\end{array}\right).
\end{array}
}
In the translation \rref{efree4}, this is
\beg{efree8}{
\exp(\tanh(\pi\epsilon)\frac{x_{-n}\tilde{x}_{-n}}{n})
\exp((-\ln\cosh\pi\epsilon)(
\frac{x_{-n}x_{n}}{n}+\frac{\tilde{x}_{-n}\tilde{x}_{n}}{n}+1))
\exp(-\tanh(\pi\epsilon)\frac{x_{n}\tilde{x}_{n}}{n}).
}
To exponentiate the middle term, we claim
\beg{efree9}{\exp(\frac{x_{-n}x_{n}}{n}z)=
:\exp\frac{x_{-n}x_{n}}{n}(e^z-1)):
}
To prove \rref{efree9}, differentiate both sides by $z$. On
the left hand side, we get
$$\frac{x_{-n}x_{n}}{n}\exp(\frac{x_{-n}x_{n}}{n}z).$$
Thus, if the derivative by $z$ of the right hand side $y$ of \rref{efree9}
is
\beg{efree10}{\frac{x_{-n}x_{n}}{n}:\exp(\frac{x_{-n}x_{n}}{n}(e^z-1)):,
}
then we have the differential equation $y^{\prime}=\frac{x_{-n}x_{n}}{n}y,$
which proves \rref{efree9} (looking also at the initial condition at
$z=0$).

Now we can calculate \rref{efree10} by moving the $x_n$ occuring before
the normal order symbol to the right. If we do this simply by changing
\rref{efree10} to normal order, we get
\beg{efree11}{:\frac{x_{-n}x_{n}}{n}\exp(\frac{x_{-n}x_{n}}{n}(e^z-1)):,
}
but if we want equality with \rref{efree10}, we must add the terms
coming from the commutator relations $[x_{n},x_{-n}]=n$, which
gives the additional term
\beg{efree12}{(e^z-1)
:\frac{x_{-n}x_{n}}{n}\exp(\frac{x_{-n}x_{n}}{n}(e^z-1)):.
}
Adding together \rref{efree11} and \rref{efree12} gives
\beg{efree13}{e^z
:\frac{x_{-n}x_{n}}{n}\exp(\frac{x_{-n}x_{n}}{n}(e^z-1)):,
}
which is the derivative by $z$ of the right hand side of \rref{efree9},
as claimed.

\vspace{3mm}
Using \rref{efree9}, \rref{efree8} becomes
\beg{efree14}{\begin{array}{l}
\Phi_n=\frac{1}{\cosh\pi\epsilon}\exp(\tanh(\pi\epsilon)
\frac{x_{-n}\tilde{x}_{-n}}{n})\\
:\exp((\frac{1}{\cosh\pi\epsilon}-1)(
\frac{x_{-n}x_{n}}{n}+\frac{\tilde{x}_{-n}\tilde{x}_{n}}{n}+1)):
\exp(-\tanh(\pi\epsilon)\frac{x_{n}\tilde{x}_{n}}{n})
\end{array}
}
which is in normal order. Let us write
\beg{eiii*}{\Phi_n=\frac{1}{\cosh\pi\epsilon}\Phi^{\prime}_{n}.}
Then the product
$$\Phi^{\prime}=\cform{\prod}{n\geq1}{}\Phi^{\prime}_{n}$$
is in normal order, and is the regularized isomorphism from
the exponentiated $\epsilon$-deformation $W_{\epsilon}$ of the conformal
field theory in vertex operator formulation on 
to the original $W$. The inverse, which goes from $W$ to
$W_{\epsilon}$, is best calculated by regularizing the
exponential of $-\phi$. We get
$$\begin{array}{l}
\exp\pi\epsilon(f-e)=\exp\pi\epsilon
\left(\begin{array}{rr}0&-1\\-1&0\end{array}\right)\\
=\left(\begin{array}{rr}\cosh\pi\epsilon&-\sinh\pi\epsilon
\\ -\sinh\pi\epsilon&\cosh\pi\epsilon\end{array}\right)\\
\left(\begin{array}{cc}1&-\tanh\pi\epsilon\\0&1\end{array}\right)
\left(\begin{array}{cc}\frac{1}{\cosh\pi\epsilon}&0\\0&\cosh\pi\epsilon
\end{array}\right)
\left(\begin{array}{cc}1&0\\-\tanh\pi\epsilon&1\end{array}\right)=\\
\frac{1}{\cosh\pi\epsilon}\exp(-\tanh(\pi\epsilon)
\frac{x_{-n}\tilde{x}_{-n}}{n})\\
:\exp((\frac{1}{\cosh\pi\epsilon}-1)(
\frac{x_{-n}x_{n}}{n}+\frac{\tilde{x}_{-n}\tilde{x}_{n}}{n}+1)):
\exp(\tanh(\pi\epsilon)\frac{x_{n}\tilde{x}_{n}}{n}).
\end{array}
$$
So expressing this as
\beg{eiii**}{\Psi_n=\frac{1}{\cosh\pi\epsilon}\Psi^{\prime}_{n},}
the product
$$\Psi^{\prime}=\cform{\prod}{n\geq 1}{}\Psi^{\prime}_{n}$$
is the regularized iso from $W$ to $W_{\epsilon}$.

Even though $\Psi^{\prime}$ and $\Phi^{\prime}$ are
only elements of $\hat{W}$, the element $u(\epsilon)=\Psi^{\prime}u$
is the regularized chiral primary field in $W_{\epsilon}$, and
can be used in a regularized version of the equation \rref{einf4}
to calculate the vacua on $V_{\epsilon}$, which will converge
on non-degenerate Segal worldsheets.

\vspace{3mm}
In this approach, however, the resulting CFT structure on $V_{\epsilon}$
remains opaque, while as it turns out, in the present case it can
be identified by another method.

\vspace{3mm}
In fact, to answer the question, we must treat precisely the
case missing in Theorem \ref{l1}, namely when the weight $0$
part of the vertex operator of the deforming field, which in
this case is determined by the momentum, doesn't vanish. 
The answer is actually known in string theory 
to correspond to constant deformation of the metric on spacetime, which 
ends up isomorphic to the original free field theory.
From the point of view of string theory, what we shall give
is a ``purely worldsheet argument'' establishing this fact.

\vspace{3mm}
Let us look first at the infinitesimal deformation of the
operator $Y(v,t,\overline{t})$ for some field $v\in V$
which is an eigenstate of momentum. We have three forms
which coincide where defined:
\beg{efree15}{Y(x_{-1}\tilde{x}_{1},z,\overline{z})Y(v,t,\overline{t})dz
d\overline{z}
}
\beg{efree16}{Y(v,t,\overline{t})Y(x_{-1}\tilde{x}_{-1},z,\overline{z})dz
\overline{z}
}
\beg{efree17}{Y(Y(x_{-1}\tilde{x}_{-1},z-t,\overline{z-t})v,t,\overline{t})
dzd\overline{z}.
}
By chiral splitting, if we assume $v$ is a monomial in the
modes, we can denote \rref{efree15}, \rref{efree16}, 
\rref{efree17} by $\eta\tilde{\eta}$ (without forming
a sum of terms). Again, integrating \rref{efree15}-\rref{efree17}
term by term $dz$, we get forms $\omega_\infty$, $\omega_0$,
$\omega_t$, respectively. Here we set
$$\int\frac{1}{z}dz=\ln z.$$
Again, these are branched forms. Selecting points $p_0,p_\infty,p_t$ on
the corresponding boundary components, we can, say, make cuts
$c_{0,t}$ and $c_{0,\infty}$ connecting the points $p_0,p_t$
and $p_0,p_\infty$. Cutting the worldsheet in this way, we obtain
well defined branches $\omega_\infty$, $\omega_0$, $\omega_t$.
To complicate things further, we have constant discrepancies
\beg{efree18}{\begin{array}{l}
C_{0t}=\omega_0-\omega_t\\
C_{0\infty}=\omega_0-\omega_\infty.
\end{array}
}
These can be calculated for example by comparing with the 
$4$ point function
\beg{efree19}{Y_+(x_{-1},z)Y(v,t)+Y(v,t)Y_-(x_{-1},z) +Y(Y_-(x_{-1},z-t)v,t)
}
where $Y_-(v,z)$ denotes the sum of the terms in $Y(v,z)$ involving negative powers
of $z$, and $Y_+(v,z)$ is the sum of the other terms. Another way to approach
this is as follows: one notices that
\beg{efree20}{\int Y(x_{-1},z)dz=\partial_{\epsilon}Y(1_{\epsilon},z)S_{-\epsilon}
|_{\epsilon=0}
}
where $S_{m}$ denotes the operator which adds $m$ to momentum. It follows
that
\beg{efree21}{\begin{array}{l}
C_{0t}=\partial_{\epsilon}(Z(x_{-1},v,z,t)S_{-\epsilon}-
Z(x_{-1},S_{\epsilon}v,z,t))|_{\epsilon=0}\\
C_{0\infty}=\partial_{\epsilon}(Z(x_{-1},v,z,t)S_{-\epsilon}-
S_{\epsilon}Z(x_{-1},v,z,t))|_{\epsilon=0}.
\end{array}
}
Now the deformation is obtained by integrating the forms
\beg{efree22}{\omega_0\tilde{\eta}}
\beg{efree23}{(\omega_t+C_{0t})\tilde{\eta}
}
\beg{efree24}{(\omega_\infty +C_{0\infty})\tilde{\eta}
}
on the boundary components around $0$, $t$ and $\infty$, and along
both sides of the cuts $c_{0t}$, $c_{0,\infty}$. To get the
integrals of the terms in \rref{efree22}-\rref{efree24} which
do not involve the discrepancy constants, we need to integrate
\beg{efree25}{(\cform{\sum}{n\neq 0}{}\frac{x_{-n}}{n}z^n+
x_0\ln z)(\cform{\sum}{m}{}x_{-m}\overline{z}^{m-1}).
}
To do this, observe that 
(pretending we work on the degenerate
worldsheet, and hence omitting scaling factors, taking curved
integrals over $||z||=1$),
\beg{efree26}{\oint\frac{\ln z}{\overline{z}}d\overline{z}=
-\oint\frac{-\ln\overline{z}}{\overline{z}}d\overline{z}=
-2\pi i\ln\overline{z}-\frac{1}{2}(2\pi i)^2
}
\beg{efree27}{\oint\ln z\cdot \overline{z}^{m-1}d\overline{z}=
-2\pi i \frac{1}{m}\overline{z}^{m}.
}
Integrating \rref{efree25}, we obtain terms
\beg{efree28}{-2\pi ix_0(\cform{\sum}{m\neq 0}{}\tilde{x}_{-m}\frac{\overline{z}^m}{m}
+\tilde{x}_0\ln\overline{z})
}
which will cancel with the integral along the cuts
(to calculate the integral over the cuts, pair points on
both sides of the cut which project to the same point in
the original worldsheet), and ``local'' terms
\beg{efree29}{\frac{2\pi i}{2}\cform{\sum}{n\neq 0}{}
\frac{x_{-n}\tilde{x}_{-n}}{n}-\frac{1}{2}(2\pi i)^2 x_0\tilde{x}_0.
}
The discrepancies play no role on the cuts (as the forms $C_{0t}\tilde{\eta}$,
$C_{0,\infty}\tilde{\eta}$ are unbranched), but using the formula
\rref{efree21}, we can compensate for the discrepancies to linear
order in $\epsilon$ by applying on each boundary component 
\beg{efree30}{S_{-2\pi i\epsilon\tilde{x}_{0}}.
}
In \rref{efree28}, however, when integrating $\tilde{\eta}$, we obtain 
also discrepancy terms conjugate to \rref{efree30}, so the correct
expression is
\beg{efree31}{S_{-2\pi i\epsilon\tilde{x}_{0}}\tilde{S}_{-2\pi i\epsilon x_{0}}.
}
The term \rref{efree31} is also ``local'' on the boundary components, so
the sum of \rref{efree29} and \rref{efree31} is the formula for the infinitesimal
iso between the free CFT and the infinitesimally deformed theory. To 
exponentiate, suppose now we are working in a $D$-dimensional
free CFT, and the deformation field is
\beg{efree32}{Mx_{-1}\tilde{x}_{-1}.
}
Then the formula for the exponentiated isomorphism multiplies left momentum by
\beg{efree33}{\exp\epsilon M
}
and right momentum by
\beg{efree34}{\exp\epsilon M^T.
}
But of course, in the free theory, the left momentum must equal
to the right momentum, so this formula works only when $M$ is a symmetric
matrix. Thus, to cover the general case, we must discuss the case when
$M$ is antisymmetric. In this case, it may seem that we obtain
indeed a different CFT which is defined in the same way as the
free CFT with the exception that the left momentum $m_L$ and right
momentum $m_R$ are related by the formula
$$m_L=Am_R$$
for some fixed orthogonal matrix $A$. As it turns out, however, this
theory is still isomorphic to the free CFT. The isomorphism replaces
the left moving oscillators $x_{i,-n}$ by their transform via the
matrix $A$ (which acts on this Heisenberg representation by transport
of structure).

\vspace{3mm}
Next, let us discuss the case of deforming gravitaitonal field of
non-zero momentum, i.e. when
\beg{efree36}{u=Mx_{-1}\tilde{x}_{-1}1_{\lambda}
}
with $\lambda\neq 0$.
Of course, in order for \rref{efree36} to be of weight $(1,1)$, 
we must have
\beg{efree37}{||\lambda||=0.
}
Clearly, then, the metric cannot be Euclidean, hence there will
be ghosts and a part of our theory doesn't apply. Note that in order
for \rref{efree36} to be primary, we also must have
\beg{efree40a}{M=\cform{\sum}{i}{}\mu_i\otimes\tilde{\mu}_i}
where
\beg{efree40}{\langle\mu_i,\lambda\rangle=\langle\tilde{\mu}_{i},\lambda\rangle
=0.
}
Despite the indefinite signature, we
still have the primary obstruction, which is
\beg{efree38}{coeff_{z^{-1}\tilde{z}^{-1}}:\cform{\sum}{m,n}{}Mx_{-m}\tilde{x}_{-n}
z^{m-1}\overline{z}^{n-1}\exp\lambda(\cform{\sum}{k\neq 0}{}
(\frac{x_{-k}}{k}z^{k}+\frac{\tilde{x}_{-k}}{k}\overline{z}^{k})):Mx_{-1}
\tilde{x}_{-1}1_{\lambda}
}
(we omit the $z^{\langle\lambda,x_0\rangle}$ term, since the power is $0$
by \rref{efree37}). In the notation \rref{efree40a}, this is
$$\begin{array}{l}
\cform{\sum}{i,j}{}(\mu_ix_0-\mu_ix_0\lambda x_{-1}\lambda x_1+
\mu_i x_{-1}\lambda x_{1}+\lambda x_{-1}\mu_i x_{1})\otimes\\
(\tilde{\mu}_j\tilde{x}_0-\tilde{\mu}_j\tilde{x}_0\lambda \tilde{x}_{-1}\lambda 
\tilde{x}_1+
\tilde{\mu}_j \tilde{x}_{-1}\lambda \tilde{x}_{1}+\lambda 
\tilde{x}_{-1}\tilde{\mu}_j \tilde{x}_{1})Mx_{-1}
\tilde{x}_{-1}1_{\lambda}
\end{array}
$$
which in the presence of \rref{efree40} reduces to the condition
\beg{efree41}{||M||^2\lambda\otimes\lambda x_{-1}\tilde{x}_{-1}=0
}
This is false unless
\beg{efreen}{||M||^2=0
}
which means that \rref{efree36} is a null state, along which the deformation
is not interesting in the sense of string theory. More generally, the distributional
form of \rref{efree41} is
\beg{efree42}{\int_{||\lambda||^2=0}\lambda\otimes\lambda||M(\lambda)||^2=0.
}
If we set
$$f(\lambda)=\delta_{||\lambda||^2=0}||M(\lambda)||^2$$
then the Fourier transform of $f$ will be a function $g$ satisfying
$$\sum\pm\frac{\partial^2g}{\partial\lambda^{2}_{i}}=0$$
where the sings correspond to the metric, which we assume is diagonal
with entries $\pm 1$. The Fourier transform of the condition \rref{efree42}
is then
\beg{efree43}{\frac{\partial^2}{\partial\lambda_i\partial\lambda_j}g=0.
}
Assuming a decay condition under which the Fourier transform makes sense,
\rref{efree43} implies $g=0$, hence \rref{efreen}, so in this case also
the obstruction is nonzero unless \rref{efree36} is a null state.

\vspace{3mm}
In this discussion, we restricted our attention to 
deforming fields of gravitational origin. It is important to note that
other choices are possible. As a very basic example, let us look
at the $1$-dimensional Euclidean model. Then there is a 
possibility of critical fields of the form
\beg{ecomm+}{a 1_{\sqrt{2}}+ b1_{-\sqrt{2}}.
}
This includes the sine-Gordon interaction \cite{sineg} when $a=b$.
(We see hyperbolic rather than
trigonometric functions because we are working in Euclidean
spacetime rather than in the time coordinate, which is
the case usually discussed.)
The primary obstruction in this case states that the 
weight $(0,0)$ descendant of \rref{ecomm+} applied to \rref{ecomm+} is $0$.
Since the descendant is
$$(4ab)x_{-1},$$
we obtain the condition $a=0$ or $b=0$. It is interesting to
note that in the case of the compactification on a circle, these
cases where investigated very successfully by Ginsparg \cite{ginsparg},
who used the obstruction to competely characterize the component
of the moduli space of $c=1$ CFT's originating from the free
Euclidean compactified free theory. The result is that only
free theories compactify at different radii, and their $\Z/2$-orbifolds
occur. 

\vspace{3mm}
There are many other possible choices of non-gravitational
deformation fields, one for each field in the physical spectrum
of the theory. We do not discuss these cases in the present paper.

\vspace{3mm}
Let us now look at the $N=1$-supersymmetric free field theory.
In this case, as pointed out above, in the NS-NS sector, 
critical gravitational fields for deformations have weight $(1/2,1/2)$.
We could also consider the NS-R and R-R sectors, where the
critical weights are $(1/2,0)$ and $(0,0)$, respectively.
These deforming fields parametrize soul directions in the
space of infinitesimal deformations. The soul parameters
$\theta$, $\tilde{\theta}$ have weights $(1/2,0)$, $(0,1/2)$,
which explains the difference of critical weights in these
sectors.

\vspace{3mm}
Let us, however, focus on the body of the space of gravitational deformations,
i.e. the NS-NS sector. Let us first look at the weight $(1/2,1/2)$ primary
field
\beg{efree46}{M\psi_{-1/2}\tilde{\psi}_{-1/2}.
}
The point is that the infinitesimal deformation is obtained by
integrating the insertion operators of 
$$G_{-1/2}\tilde{G}_{-1/2}M\psi_{-1/2}\tilde{\psi}_{-1/2}=
Mx_{-1}\tilde{x}_{-1}.$$
Therefore, \rref{efree46} behaves exactly the same was as deformation
along the field \rref{efree32} in the bosonic case. Again, if $M$ is
a symmetric matrix, exponentiating the deformation leads to a theory
isomorphic via scaling the momenta, while if $M$ is antisymmetric,
the isomorphism involves transforming the left moving modes by the
orthogonal matrix $\exp(M)$.

\vspace{3mm}
In the case of momentum $\lambda\neq 0$, we again have indefinite
signature, and the field
\beg{efree48}{u=M\psi_{-1/2}\tilde{\psi}_{-1/2}1_{\lambda}.
}
Once again, for \rref{efree48} to be primary, we must have
\rref{efree40a}, \rref{efree40}.
Moreover, again the actual infinitesimal deformation is
got by applying the insertion operators of 
$G_{-1/2}\tilde{G}_{-1/2}u$, so the treatment is exactly the
same as deformation along the field \rref{efree36} in the bosonic
case. Again, we discover that under a suitable decay condition, 
the obstruction is always nonzero for gravitational deformations
of non-zero momentum with suitable decay conditions.

\vspace{3mm}
It is worth noting that in both the bosonic and supersymmetric
cases, one can apply the same analysis to free field theories
compactified on a torus. In this case, however, scaling momenta
changes the geometry of the torus, so using deformation fields
of $0$ momentum, we find exponential deformations which change
(constantly) the metric on the torus. This seems to confirm, in
the restricted sense investigated here, a conjecture stated
in \cite{scft}.

\vspace{3mm}
\noindent 
{\bf Remark:}
Since one can consider Calabi-Yau manifolds which are tori,
one sees that there should also exist an $N=2$-supersymmetric
version of the free field theory compactified on a torus. 
(It is in fact not difficult to construct such model directly,
it is a standard construction.) Now since we are in 
the Calabi-Yau case, marginal $cc$ fields should correspond
to deformations of complex structure, and marginal $ac$ fields
should correspond to deformations of K\"{a}hler metric in this
case. 

But on the other hand, we already identified gravitational
fields which should be the sources of such deformations.
Additionally, deformations in those direction require 
regularization of the deformation parameter, and hence
cannot satisfy the conclusion of Theorem \ref{t2}.

This is explained by observing that we must be careful with reality.
The gravitational fields we considered are in fact real,
but neither chiral nor antichiral primary in either the left or
the right moving sector. By contrast, chiral primary fields (
or antiprimary) fields are not real. This is due to the fact
that $G^{+}_{-3/2}$ and $G^{-}_{-3/2}$ are not real
in the $N=2$ superconformal algebra, but are
in fact complex conjugate to each other. Therefore, to get
to the real gravitational fields, we must take real parts,
or in other words linear combinations of chiral and antichiral
primaries, resulting in the need for regularization.

\vspace{3mm}
It is in fact a fun exercise to calculate explicitly
how our higher $N=2$ obstruction theory operates in this
case. Let us consider the $N=2$-supersymmetric free
field theory, since the compactification behaves analogously. 
The minimum number of dimensions for $N=2$ supersymmetry
is $2$. Let us denote the bosonic fields by $x,y$ and
their fermionic superpartners by $\xi,\psi$. Then the
$0$-momentum summand of the state space (NS sector)
is (a Hilbert completion of)
$$Sym(x_n,y_n|n<0)\otimes \Lambda(\xi_r,\psi_r|
r<0,r\in\Z+\frac{1}{2}).$$
The ``body'' parts of the bosonic and fermionic vertex operators
are given by the usual formulas
$$\begin{array}{ll}
Y(x_{-1},z)=\sum x_{-n}z^{n-1}, &
Y(y_{-1},z)=\sum y_{-n}z^{n-1}\\
Y(\xi_{-1/2},z)=\sum \xi_{-s}z^{n-s-1/2}&
Y(\psi_{-1/2},z)=\sum \psi_{-s}z^{n-s-1/2},
\end{array}$$
$$\begin{array}{l}
[\xi_r,\xi_{-r}]= [\psi_r,\psi_{-r}]=1\\{}
[x_n,x_{-n}]= [y_n,y_{-n}]=n.
\end{array}$$
We have, say,
$$\begin{array}{l}
G^{1}_{-3/2}=\xi_{-1/2}x_{-1}+\psi_{-1/2}y_{-1}\\
G^{2}_{-3/2}=\xi_{-1/2}y_{1}-\psi_{-1/2}x_{-1}.
\end{array}
$$
As usual,
$$G^{\pm}_{-3/2}=\frac{1}{\sqrt{2}}(G^{1}_{-3/2}\pm i G^{2}_{-3/2}).$$
With these conventions, we have a critical chiral primary
\beg{efun0}{u=\xi_{-1/2}-i\psi_{-1/2}
}
(and its complex conjugate critical antichiral primary).
We then see that for a non-ero coefficient $C$,
\beg{efun1}{CG^{-}_{-1/2}u=x_{-1}-iy_{-1}.
}
We now notice that formulas analogous to \rref{efree4} etc.
apply to \rref{efun1}, but the $-1$ summans of $h$ will
appear with opposite signs for the real and imaginary
summands, so it will cancel out, so the regularization
\rref{eiii*}, \rref{eiii**} are not needed, as expected. 

Next, let us study the formula \rref{ecorft2a}.
The key observation here is that
we have the combinatorial identity
\beg{efun2}{\frac{1}{n_1...n_k}=
\cform{\sum}{\sigma}{}\frac{1}{(n_{\sigma(1)}+...n_{\sigma(k)})
(n_{\sigma(1)}+...n_{\sigma(k-1)})...n_{\sigma(1)}}
}
where the sum on the right is over all permutations on the
set $\{1,...,k\}$. Now in the present case, we have the
infinitesimal iso on the $0$-momentum part, up to non-zero 
coefficient,
\beg{efun3}{\phi=
\sum \frac{(x_n-iy_n)(\tilde{x}_{n}+i\tilde{y}_{n})}{n}
}
and in the absence of regularization, the expansion of the
exponentiated isomorphism on the $0$-momentum parts is
simply
$$\exp(\phi\epsilon).$$
(The $+$ sign in the $\tilde{\:}$'s is caused by the fact that
we are in the complex conjugate Hilbert space.)
Applying this to \rref{efun0}, we see that we have formulas
analogous to \rref{efree8}-\rref{efree14},
and applying the exponentiated iso to \rref{efun0},
all the terms in normal order involving $x_{>0}$, $y_{>0}$
will vanish, so we end up with
$$
\cform{\prod}{n<0}{}\exp(D\frac{(x_n-iy_n)(\tilde{x}_n+i\tilde{y}_n)}{
n})u$$
for some non-zero coefficient $D$. Applying \rref{efun2},
we get \rref{ecorft2a}.

Finally, the obstruction in chiral form
$$\langle u^{*}(0),(G^{-}_{-1/2}u)(z_k),...,(G^{-}_{-1/2}u)(z_1),u(0)
\rangle$$
must vanish identically. To see this, we simply observe
\rref{efun0}, \rref{efun1} that in the present case,
$u$ is in the coset model with respect to $G^{-}_{-1/2}u$
(see the discussion below formula \rref{eg66} below). 
Thus, in the $N=2$-free field theory, the obstruction
theory works as expected, and in the case discussed, the 
obstructions vanish. It is worth noting that in $2n$-dimensional
$N=2$-free field theory, we thus have an $n^2$-dimensional 
space of $cc+aa$ real fields, and an $n^2$-dimensional space of
real $ca+ac$ 
fields, and although regularization occurs, there is no obstruction
to exponentiating the deformation by turning on any linear combination
of those fields. For a free $N=2$-theory compactified on an
$n$-dimensional abelian variety, this precisely recovers the deformations
in the corresponding component of the moduli space of Calabi-Yau varieties.

However, other deformations exist. For an interesting calculation
of deformations of the $N=2$-free field theory in ``sine-Gordon''
directions, see \cite{cgep}.

\vspace{10mm}

\section{The Gepner model of the Fermat quintic}

\label{s1}

\vspace{3mm}
The finite weight states of one chirality (say, left moving) of
the Gepner model of the Fermat quintic are embedded in the
$5$-fold tensor product of the $N=2$-supersymmetric minimal model
of central charge $9/5$ \cite{gepner,gepner1, gp}. More precisely,
the Gepner model is an orbifold construction. This construction
has two versions. In \cite{gepner, gepner1, gp}, one is interested
in actual string theories, so the $5$-fold
tensor product of central charge $9$ of $N=2$ minimal models
is tensored with a free supersymmetric CFT on $4$ Minkowski coordinates.
This is then viewed in lightcone gauge, so in effect, one
tensors with a $2$-dimensional supersymmetric Euclidean
free CFT, resulting in $N=2$-supersymmetric CFT of central
charge $12$. Finally, one performs an orbifolding/GSO projection
to give a candidate for a theory for which both modularity and
spacetime SUSY can be verified. 

\vspace{3mm}
It is also possible to create an orbifold theory of central charge
$9$ which is the candidate of the non-linear $\sigma$-model
itself, without the spacetime coordinates. (The spacetime
coordinates can be added to this construction
and usual GSO projection performed
if one is interested in the corresponding string
theory.) 

\vspace{3mm}
The essence of this
construction not involving the spacetime coordinates
is formula (2.10) of \cite{gvw}. In the case
of the level $3$ $N=2$-minimal model, the orbifold construction
is with respect to the $\Z/5$-action diagonal which acts on the 
eigenstates of $J_0$-eigenvalue (=``$U(1)$-charge'')
$j/5 $ by $e^{2\pi ij/5}$.
As we shall review, the NS part of the 
level $3$ $N=2$ minimal model has 
two sectors of $U(1)$-charge $j/5$, which we will for the moment
ad hoc denote $H_{j/5}$ and $H^{\prime}_{j/5}$ for $j\in \Z/5\Z$.
In the FF realization (see below), these sectors correspond
to $\ell=0$, $\ell=1$, respectively. Then the NS-NS sector
of the $5$-fold tensor product of minimal model has the
form
\beg{egepp1}{
\cform{\sum}{(i_k)}{} \cform{\hat{\bigotimes}}{k=1}{5}
(H_{i_k/5}\hat{\otimes} H^{*}_{i_k/5}\oplus H^{\prime}_{i_k/5}\hat{\otimes} 
H^{\prime*}_{i_k/5}).
}
The corresponding sector the orbifold construction (formula (2.10)
of \cite{gvw}) of the orbifold construction has the form
\beg{egepp2}{
\cform{\sum}{(i_k):\sum i_k\in 5\Z}{} \cform{\sum}{j\in \Z/5\Z}{}
\cform{\hat{\bigotimes}}{k=1}{5} 
(H_{i_k/5}\hat{\otimes} H^*_{(j+i_k)/5}\oplus H^{\prime}_{i_k/5}\hat{\otimes} 
H^{\prime*}_{(j+i_k)/5}).
}
Mathematically speaking, this orbifold can be constructed by noting
that, ignoring for the moment supersymmetry, the $N=2$-minimal
model is a tensor product of the parafermion theory of the same
level and a lattice theory (see \cite{greene} and also below).
The orbifold construction does not affect the parafermionic
factor, and on the lattice coordinate, which in this case
does not possess a non-zero $\Z/2$-valued form, and hence
physically models a free theory compactified on a torus, the
orbifold simply means replacing the torus by its factor by
the free action of the diagonal $\Z/5$ translation group,
which is represented by another lattice theory. On this construction,
$N=2$ supersymmetry is then easily restored using the same
formulas as in \rref{egepp1}, since the $U(1)$-charge of the
$G$'s is integral. 

\vspace{3mm}
The calculations in this and the next Section proceed entirely
in the orbifold \rref{egepp2}, and hence can be derived
from the structure of the level $3$ $N=2$-minimal model. It
should be pointed out that a mathematical
approach to the fusion rules of the $N=2$ minimal models was 
given in \cite{huang1}. We shall use the Coulomb gas realization
of the $N=2$-minimal model, cf. \cite{gh}, \cite{mus}. Let us
restrict attention to the NS sector. Then, essentially, the 
left moving sector of the minimal model is a subquotient of the 
lattice theory where the lattice is $3$-dimensional, and spanned by 
$$(\frac{3}{\sqrt{15}},0,0),(\frac{1}{\sqrt{15}}, 
\frac{i}{2}\sqrt{\frac{2}{5}},\frac{i}{2}\sqrt{\frac{2}{3}}),
(\frac{2}{\sqrt{15}},0,i\sqrt{\frac{2}{3}}).$$
We will adopt the convention that we shall abbreviate
$$(k,\ell,m)_{MM}=(k,\ell,m)$$
for the lattice label
$$(\frac{k}{\sqrt{15}},\frac{\ell i}{2}\sqrt{\frac{2}{5}},\frac{mi}{2}
\sqrt{\frac{2}{3}}).
$$
We shall also write
$$(\ell,m)_{MM} =(m,\ell,m)_{MM}.$$
Call the oscillator corresponding to the $j$'th
coordinate $x_{j,m}$, $j=0,1,2$. Then the conformal vector is
\beg{eg7}{\frac{1}{2}x_{0,-1}^{2}-\frac{1}{2}x_{1,-1}^{2}
+\frac{i}{2}\sqrt{\frac{2}{5}}x_{1,-2}+\frac{1}{2}x_{2,-1}^{2}.
}
The superconformal algebra is generated by
\beg{eg8}{{
\begin{array}{l}
G^{+}_{-3/2}=i\sqrt{\frac{1}{2}}(x_{2,-1}-\sqrt{\frac{5}{3}}x_{1,-1})
1_{(\frac{5}{\sqrt{15}},0,i\sqrt{\frac{2}{3}})}\\
G^{-}_{-3/2}=-i\sqrt{\frac{1}{2}}(x_{2,-1}+\sqrt{\frac{5}{3}}x_{1,-1})
1_{(-\frac{5}{\sqrt{15}},0,-i\sqrt{\frac{2}{3}})}.
\end{array}
}}
For future reference, we will sometimes use the notation
$$(a,b,c)x_{n}=ax_{0,n} +bx_{1,n}+cx_{2,n}$$
and also sometimes abbreviate
$$(b,c)x_{n}=(0,b,c)x_{n}.$$
The module labels are realized by labels
\beg{eg9}{(\ell,m)=1_{(\frac{m}{\sqrt{15}},-\frac{i\ell}{2}\sqrt{
\frac{2}{3}},\frac{im}{2}\sqrt{\frac{2}{3}})},
}
\beg{eg9a}{0\leq\ell\leq 3,\; m=-\ell,-\ell+2,...,\ell-2,\ell.
}
It is obvious that to stay within the range \rref{eg9a}, we
must understand the fusion rules and how they are applied.
The basic principle is that labels are indentified as follows:
No identifications are imposed on the $0$'th lattice coordinate.
This means that upon any identification, the $0$'th
coordinate must be the same for the labels identified. Therefore,
the identification is governed by the $1$st and $2$nd coordinates,
which give the Coulomb gas (=Feigin-Fuchs)
realization of the corresponding
parafermionic theory (the ``$3$ state Potts model'').
The key point here are the parafermionic currents
\beg{eg10}{{
\begin{array}{l}
\psi_{1,-2/3}=i\sqrt{\frac{1}{2}}(x_{2,-1}-x_{1,-1}\sqrt{\frac{5}{3}})
1_{(0,i\sqrt{\frac{2}{3}})_{PF}}\\
\psi_{1,-2/3}^{+}=-i\sqrt{\frac{1}{2}}(x_{2,-1}+x_{1,-1}\sqrt{\frac{5}{3}})
1_{(0,-i\sqrt{\frac{2}{3}})_{PF}}
\end{array}
}}
(the $0$'the coordinate is omitted). Clearly, the parafermionic currents
act on the labels by
\beg{11}{{
\begin{array}{l}
\psi_{1,-2/3}:(\ell,m)_{PF}\mapsto (\ell,m+2)_{PF}\\
\psi_{1,-2/3}^{+}:(\ell,m)_{PF}\mapsto (\ell,m-2)_{PF}.
\end{array}
}}
The lattice labels $(\ell,m)_{PF}$ allowed are those which have 
non-negative weight. This condition coincides with \rref{eg9a}.
Now we impose the identification for
parafermionic labels:
$$(\ell,m)_{PF}=(3-\ell,m-3)_{PF}.$$
This implies
\beg{eg12}{{
\begin{array}{l}
(1,-1)_{PF}\sim(2,2)_{PF}\\
(1,1)_{PF}\sim (2,-2)_{PF}\\
(0,0)_{PF}\sim (3,-3)_{PF}\sim (3,3)_{PF}.
\end{array}
}}

\vspace{3mm}
Now in the Gepner model corresponding to the
quintic, the $(cc)$-fields allowed are
\beg{eg15}{((3,2,0,0,0)_{L},(3,2,0,0,0)_R),
}
\beg{eg16}{((3,1,1,0,0)_{L},(3,1,1,0,0)_R),
}
\beg{eg17}{((2,2,1,0,0)_{L},(2,2,1,0,0)_R),
}
\beg{eg18}{((2,1,1,1,0)_{L},(2,1,1,1,0)_R),
}
\beg{eg19}{((1,1,1,1,1)_{L},(1,1,1,1,1)_R),
}
and the $(ac)$-field allowed is
\beg{eg20}{((-1,-1,-1,-1,-1)_{L},(1,1,1,1,1)_R).
}
Here we wrote $\ell$ for $1_{(\ell,\ell)_{MM}}$,
$(\ell=0,...,3)$, which is a chiral primary in the 
$N=2$ minimal model of weight $\ell/10$, and
$-\ell$ for $1_{(\ell,-\ell)_{MM}}$, which is 
antichiral primary of weight $\ell/10$. 
The tuple notation in \rref{eg15}-\rref{eg20}
really means tensor product.
We omit permutations of the fields \rref{eg15}-\rref{eg18},
so counting all permutations, there are $101$ fields
\rref{eg15}-\rref{eg19}.

\vspace{3mm}
We will need an understanding of the fusion rules
in the level $3$ Potts model and $N=2$-supersymmetric
minimal model of central
charge $9/5$. In the level $3$ Potts model, we have
$6$ labels
\beg{eF1}{(0,0)_{PF},\; (3,1)_{PF},\; (3,-1)_{PF},
}
\beg{eF2}{(1,1)_{PF},\; (1,-1)_{PF},\; (2,0)_{PF}.
}
This can be described as follows: the labels
\rref{eF1} have the same fusion rules as the lattice
$L=\langle i\sqrt{6}\rangle\subset \C$, i.e.
\beg{eF3}{L^{\prime}/L
}
where $L^{\prime}$ is the dual lattice (into which $L$
is embedded using the standard quadratic form on $\C$).
This dual lattice is $\langle \frac{i}{2}\sqrt{\frac{2}{3}}\rangle$,
and the fusion rule is ``abelian'', which means that
the product of labels has only one possible label as outcome,
and is described by the product in $L^\prime/L$.
The label $\pm \frac{i}{2}\sqrt{\frac{2}{3}}$
corresponds to $(0,\pm2)_{PF}\sim (3,\mp1)_{PF}$.

Next, the product of $(2,0)_{PF}$ with
$(3,\mp 1)_{PF}$ has only one possible outcome,
$(2,\pm 2)_{PF}=(1,\mp 1)_{PF}$. The product of
$(2,0)_{PF}$ with itself has two possible outcomes,
$(2,0)_{PF}$ and $(0,0)_{PF}$. All other products
are determined by commutativity, associativity and
unitality of fusion rules. 

The result can be summarized as follows: We call \rref{eF1}
level $0,3$ labels and \rref{eF2} level $1,2$ labels.
Every level $1,2$ label has a corresponding label
of level $0,3$. The correspondence is
\beg{eF4}{
\begin{array}{lll}
(0,0)_{PF}&\leftrightarrow&(2,0)_{PF}\\
(3,1)_{PF}&\leftrightarrow&(1,1)_{PF}\\
(3,-1)_{PF}&\leftrightarrow&(1,-1)_{PF}.
\end{array}
}
As described above, the fusion rules on level $0,3$ are
determined by the lattice theory of $L$. Additionally, 
multiplication preserves the correspondence \rref{eF4},
while the level of the product is restricted 
only by requiring that any level
added to level $0,3$ is the original level. 

To put it in another way still, the Verlinde algebra is
\beg{eF5}{\Z[\zeta]/(\zeta^3-1)\otimes \Z[\epsilon]/(\epsilon^2-
\epsilon-1)
}
where $\zeta=(3,1)_{PF}$ and $\epsilon=(2,0)_{PF}$.

In the $N=2$ supersymmetric minimal model (MM) case, we allow 
labels 
\beg{eF*}{(3k+m,\ell,m)_{MM}} 
where $(\ell,m)$ is a PF label,
$k\in \Z$. Two labels \rref{eF*} are identified subject to identifications
of PF labels, and also
\beg{eF6}{(j,\ell,m)_{MM}\sim (j+15,\ell,m)_{MM},
} 
and, as a result of SUSY, 
\beg{eF7}{(j,\ell,m)_{MM}\sim (j-5,\ell,m-2)_{MM}.
}
(By $\sim$ we mean that the labels (i.e. VA modules) are identified, but
we do not imply that the states involved actually coincide; in
the case \rref{eF7}, they have different weights.)
Recalling again that we abbreviate $(m,\ell,m)_{MM}$ as $(\ell,m)_{MM}$,
we get the following labels for the $c=9/5$ $N=2$ SUSY MM:
\beg{eF8}{\begin{array}{lll}
(0,0)_{MM}&\leftrightarrow&(2,0)_{MM}\\
(3,3)_{MM}&\leftrightarrow&(2,-2)_{MM}\\
(3,1)_{MM}&\leftrightarrow&(1,1)_{MM}\\
(3,-1)_{MM}&\leftrightarrow&(1,-1)_{MM}\\
(3,-3)_{MM}&\leftrightarrow&(2,2)_{MM}.
\end{array}
}
Again, the left column \rref{eF8} represents $0,3$ level labels, the
right column represents level $1,2$ labels. The left column labels
multiply as the labels of the lattice super-CFT corresponding 
to the lattice $\Lambda$ in $\C\oplus\C$ spanned by
\beg{eF9}{(\sqrt{15},0),\; (\frac{5}{\sqrt{15}},i\sqrt{\frac{2}{3}})
}
(recall that a super-CFT can be assigned to a lattice with integral 
quadratic form; the quadratic form on $\C\oplus\C$ is the standard
one, the complexification of the Euclidean inner product). The dual
lattice of \rref{eF9} is spanned by
\beg{eF10}{(\frac{3}{\sqrt{15}},0),\; (\frac{5}{\sqrt{15}},i\sqrt{\frac{2}{3}}),
}
which correspond to the labels $(3,0,0)_{MM}$, $(5,3,-1)_{MM}$, respectively.
We see that
\beg{eF11}{\Lambda^{\prime}/\Lambda\cong \Z/5.
}
In \rref{eF8}, the rows (counted from top to bottom as $0,...,4$)
match the corresponding residue class \rref{eF11}. The fusion rules
for $(2,0)_{MM}$, $(0,0)_{MM}$ are the same as in the PF case.
Hence, again, multiplication of labels preserves the rows \rref{eF8},
and the Verlinde algebra is isomorphic to
\beg{eF12}{\Z[\eta]/(\eta^5-1)\otimes \Z[\epsilon]/(\epsilon^2-
\epsilon-1)
}
where $\eta$ is $(3,3)_{MM}$.

\vspace{3mm}
\noindent
{\bf Remark:} As remarked in Section \ref{sexp}, the positive 
definiteness of the modular functor, which is crucial for
our theory to work, is a requirement for a physical CFT.
It is interesting to note, however, that if we do not include
this requirement, other possible choices of real structure
are possible on the modular functor: The 
Verlinde algebra  of a lattice modular functor with
another modular functor $M$ with two labels $1$ and $\epsilon$,
and Verlinde algebras \rref{eF5}, \rref{eF12} are tensor
products of lattice Verlinde algebras and the algebra
$$\Z[\epsilon]/(\epsilon^2-\epsilon-1).$$
The real structure of this last modular functor can be changed
by multiplying by $-1$ the complex conjugation in $M_\Sigma$
for a worldsheet $\Sigma$ precisely when $\Sigma$ has an odd
number of boundary components labelled on level $1$, $2$. The resulting
modular functor of this operation is not positive-definite.

\vspace{3mm}
Let us now discuss the question of vertex operators in the PF realization
of the minimal model. Clearly, since the $0$'th coordinate acts as a lattice
coordinate and is not involved in renaming, it suffices the question for the
parafermions. Now in the Feigin-Fuchs realization of the level $3$ PF model,
any state can be written as
\beg{eFst}{u 1_{\lambda}}
where $\lambda$ is one of the labels \rref{eg9a} and $u$ is
a state of the Heisenberg representation of the Heisenberg
algebra generated by $x_{i,m}$, $i=1,2$, $m\neq 0$. The situation
is however further complicated by the fact that not all 
Heisenberg states $u$ are allowed for a given label $\lambda$. 
We shall call the states which are in the image of 
the embedding admissible. For example, since the $\lambda=0$
part of the PF model is isomorphic to the 
coset model $SU(2)/S^1$ of the same level, states 
\beg{eFst1}{(a,b)x_{-1}(0,0)_{PF}}
are not admissible for $(a,b)\neq (0,0)$. 
One can show that admissible states are exactly those which are
generated from the ground states \rref{eg9a} by vertex operators
and PF currents. 
Because not all states
are admissible, however, there are also states whose vertex operators
are $0$ on admissible states. Let us call them null states. For example, since 
\rref{eFst1} is not admissible, it follows that
\beg{eFst2}{(a,b)x_{-1}(3,3)_{PF},}
which is easily seen to be admissible for any choice of $(a,b)$, is null.

Determining explicitly which states are admissible and which are
null is extremely tricky (cf. \cite{gh}). Fortunately, we don't need to
address the question for our purposes. This is because we will only
deal with states which are explicitly generated by the primary fields,
and hence automatically admissible; because of this, we can ignore
null states, which do not affect correlation functions of admissible
states. 

On the other hand, we do need an explicit formula for vertex operators.
One method for obtaining vertex operators is as follows. 
We may rename fields using the identifications \rref{eg12} 
and also PF currents: a PF current applied to a renamed field must
be equal to the same current applied to the original field. Note that
this way we may get Heisenberg states above labels which fail to
satisfy \rref{eg9a}. Such states are also admissible, even though
the corresponding ``ground states'' (which have the same name
as the label) are not. Now if we have two admissible states
$$u_i 1_{(\ell_i, m_i)},\; i=1,2$$
where $0\leq \ell_i\leq 3$ and $\ell_1+\ell_2\leq 3$, then
the lattice vertex operator
\beg{eFst10}{(u_1 1_{(\ell_1,m_1)})(z)u_21_{(\ell_2,m_2)}}
always satisfies our fusion rules, and (up to scalar multiple
constant on each module)
is a correct vertex operator of the PF theory. This is easily seen
simply by the fact that \rref{eFst10} intertwines correctly
with module vertex operators (which are also lattice operators).

While in our examples, it will suffice to always consider operators
obtained in the form \rref{eFst10}, it is important to realize that
they do not describe the PF vertex operators completely. The problem is
that when we want to iterate vertex operators, we would have to keep
renaming states. But when two ground states $1_{\lambda}$,
$1_{\mu}$ are identified
via the formula \rref{eg12}, it does not follow that we would have
\beg{eFst+}{u 1_{\lambda} =u 1_{\mu}
}
for every Heisenberg state $u$. On the contrary, we saw for example
that \rref{eFst1} is inadmissible, while \rref{eFst2} is null. One also
notes that one has for example the identification
\beg{eFst+1}{\lambda x_{-1}1_{\lambda}=L_{-1}1_{\lambda}=
L_{-1}1_{\mu} = \mu x_{-1}1_{\mu},
}
which is not of the form \rref{eFst+}. 

Because of this, to describe completely the full force of the
PF theory, one needs another
device for obtaining vertex operators (although we will not
need this in the present paper). Briefly, it is shown in \cite{gh}
that up to scalar multiple, any vertex operator 
$$u(z)v=Y(u,z)(v)$$
where $u,v$ are admissible states can be written as
\beg{eFst++}{\oint...\oint (a_k x_{-1}1_{(0,-2)})(t_k)...(a_1 x_{-1}1_{(0,-2)}(t_1)
u(z)v dt_1...dt_k
}
where the operators in the argument \rref{eFst++} are lattice vertex
operators and the number $k$ is selected to conform with the given fusion
rule. While it is easy to show that operators of the form
\rref{eFst++} are correct vertex operators on admissible states (again
up to scalar multiple constant on each irreducible module),
as the ``screening operators''
$$ax_{-1}1_{(0,-2)}$$
commute with PF currents, selecting the bounds of integration (``contours'')
is much more tricky. Despite the notation, it is not correct to imagine
these as integrals over closed curves, at least not in general. 
One approach which works is to bring the argument of \rref{eFst++} to
normal order, which expands it as an infinite sum of terms of the form
\beg{eFst+++}{\prod (t_i-t_j)^{\alpha_{ij}}t_k^{\beta_k}
}
(where we put $t_0=z$) with coefficients which are lattice vertex operators. 
Then to integrate \rref{eFst+++}, for $\alpha_{ij},\beta_k>0$, we may
simply integrate $t_i$ from $0$ to $t_{i-1}$, and define the integral
by analytic continuation in the variables $\alpha_{ij}$, $\beta_k$ otherwise.

The functions obtained in this way are generalized hypergeometric functions, 
and fail for example the assumptions of Theorem \ref{l1} (see
Remark 2 after the Theorem). The explanation is in the fact that,
as we already saw, the fusion rules are not ``abelian'' in this case.

\vspace{10mm}

\section{The Gepner model: the obstruction}
\label{sexam}

\vspace{3mm}
We will now show that for the Gepner model of the Fermat quintic, 
the function \rref{ecorelii} may not vanish for the deforming field
\rref{eg15}. This means, not all perturbative
deformations corresponding to marginal fields exist in this case. 
We emphasize that our result applies to deformations of the CFT
itself (of central charge $9$). A different approach is possible
by embedding the model to string theory, and investigating the
deformations in that setting (cf. \cite{dkl}). Our results
do not automatically apply to deformations in that setting.

\vspace{3mm}
We will consider 
$$v=u=(3,3,3)\otimes (2,2,2).$$
(In the remaining three coordinates, we will always put the vacuum, so
we will omit them from our notation.)
First note that by Theorem \rref{t2}, this is actually the only
relevant case \rref{ecorelii}, since the only other chiral primary
field of weight $1/2$ with only two non-vacuum coordinates is
$(2,2,2)\otimes (3,3,3)$, which cannot correlate with the right hand
side of \rref{ecorelii}, whose first coordinate is on level $0,3$. 
In any case, we will show therefore that the Gepner model has
an obstruction against continuous perturbative deformation along
the field \rref{eg15} in the moduli space of exact conformal field theories. 

\vspace{3mm}
Now the chiral correlation function \rref{ecorelii} is a complicated
multivalued function because of the integrals \rref{eFst+++}, which
are generalized hypergeometric functions. As remarked above, the modular functor
has a canonical flat connection on the space of degenerate worldsheets
whose boundary components are shifts of the unit circle with the
identity parametrization. The flat connection comes from the fact that 
these degenerate worldsheets are related to each other by applying
$\exp (zL_{-1})$ to their boundary components. This is why we
can speak of analytic continuation of a branch of the correlation
function corresponding to a particular fusion rule. It can
further be shown (although we do not need to use that result here)
that the continuations of the correlation function corresponding to
any one particular fusion rules generate the whole correlation function
(i.e. the whole modular functor is generated by any one non-zero section).

\vspace{3mm}
Let us now investigate which number $m$ we need in \rref{ecorelii}.
In our case, we have
\beg{eexam+}{G_{-1/2}^{-}(u)=G_{-1/2}^{-}(3,3,3)\otimes (2,2,2)-
(3,3,3)\otimes G^{-}_{-1/2}(2,2,2).
}
(The sign will be justified later; it is not needed at this point.)
The first summand \rref{eexam+} has $x_{0,0}$-charge
$(-2/\sqrt{15},2/\sqrt{15})$, the second has $x_{0,0}$-charge
$(3/\sqrt{15},-3/\sqrt{15})$. Thus, the charges can add up to $0$
only if $m$ is a multiple of $5$. The smallest possible obstruction
is therefore for $m=5$, in which case \rref{ecorelii} is
a $7$ point function. Let us focus on this
case. This function however is too big to calculate completely. Because
of this, we use the following trick. 

First, it is equivalent to consider the question
of vanishing of the function
\beg{eexam1}{\langle 1|(G^{-1}_{-1/2}u)(z_5)...(G^{-1}_{-1/2}u)(z_1)
u(z_0)\overline{u}(t)\rangle.
}
Now by the OPE, it is possible to transform any correlation function
of the form
\beg{eexam2}{\langle...|...v(z)w(t)...\rangle
}
to the correlation function
\beg{eexam3}{\langle...|...(v_nw)(t)...\rangle
}
(all other entries are the same). More precisely, \rref{eexam2} is
expanded, in a certain range and choice of branch, into a series in
$z-t$ with coefficients \rref{eexam3} for values of $n$ belonging
to a coset $\Q/\Z$. By the above argument, therefore, the function
\rref{eexam2} vanishes if and only if the function \rref{eexam3}
vanishes for all possible choices of $n$ associated with one fixed
choice of fusion rule. 

In the case of \rref{eexam1}, we shall divide the fields on the
right hand side into two sets $G_x$, $G_y$ containing two copies 
of $G^{-}_{-1/2}$ each, and a set $G_z$ containing the remaining
three fields $u, \overline{u}$ and $G^{-}_{-1/2}$. Each set 
$G_x$, $G_y$, $G_z$ will be reduced to a single field using
the transition from \rref{eexam2} to \rref{eexam3} (twice in the
case of $G_z$). To simplify notation (eliminating the subscripts),
we will denote the fields resulting from $G_x$, $G_y$, $G_z$
by $a(x)$, $b(y)$, $c(z)$, respectively. Thus, $x,y,z$ are
appropriate choices among the variables $z_i,t$, depending how the
transition from \rref{eexam2} to \rref{eexam3} is applied.

This reduces the correlation function \rref{eexam1} to 
\beg{eexam4}{\langle 1| a(x) b(y) c(z)\rangle.
}
Most crucially, however, we make the following simplification: 
We shall choose the fusion rules in such a way that the fields 
$a,b,c$ are level $0,3$ in the Feigin-Fuchs realization, and
at most one of the charges will be $3$ 
(in each coordinate). Then, \rref{eexam4} is just a lattice
correlation function, for the computation of which we have an
algorithm. 

\vspace{3mm}
To make the calculation correctly, we must keep careful track
of signs. When taking a tensor product of super-CFT's,
one must add appropriate signs analogous to the Koszul-Milnor
signs in algebraic topology. Now a modular functor of a 
super-CFT decomposes into an even part and an odd part.
Additionally, more than one choice of this decomposition
may be possible for the same theory, depending on which
bottom states of irreducible modules are chosen as even or odd. 
The sign of a fusion rule is then determined by whether 
composition along the pair of pants with given labels preserves
parity of states or not. 
Mathematically, this phenomenon was noticed by Deligne in the
case of the determinant line (cf. \cite{spin}). (Deligne also
noticed that in some cases no consistent choice of signs is
possible and a more refined formalism is needed; a single fermion
of central charge $1/2$
is an example; this is also discussed in \cite{spin}. However, this
will not be needed here.)

In the case of the $N=2$-minimal model, there is a choice of parities
of ground states of irreducible modules which make the whole modular
functor (all the fusion rules) even: simply choose the parity
of $(k,\ell,m)$ to be $k\mod 2$. We easily see that this is
compatible with supersymmetry. 

Now in this case of completely even modular functor, the signs
simplify, and we put
\beg{eexam4a}{Y(u\otimes v,z)(r\otimes s)=
(-1)^{\pi(u)\pi(v)}Y(u,z)r\otimes Y(v,z)s
}
(where $\pi(u)$ means the parity of $u$).
Regarding supersymmetry (if present), an element $H$ of the superconformal
algebra also acts on a tensor product by
\beg{eexam5}{H(u\otimes v)= Hu\otimes v +(-1)^{\pi(H)\pi(u)}u\otimes Hv,
}
in particular
\beg{eexam6}{G^{-}_{-1/2}(u\otimes v)=
(G^{-}_{-1/2}u)\otimes v +(-1)^{\pi(u)}u\otimes (G^{-}_{-1/2}v).
}
We see that because of \rref{eexam6}, the fields $a,b,c$ may have the
form of sum of several terms.

\vspace{3mm}
\noindent
{\bf Example 1:}
Recall that the inner product (more precisely symmetric bilinear
form) of labels considered as lattice points is
\beg{eexam7}{\langle(r_1,s_1,t_1),(r_2,s_2,t_2)\rangle=
\frac{r_1 r_2}{15}+\frac{s_1 s_2}{10}-\frac{t_1 t_2}{6}.
}
Recall also (from the definition of energy-momentum tensor)
that weight of the label ground states is calculated by
\beg{eexam8}{w(r,s,t)_{MM}=\frac{r^2}{30}+w(s,t)_{PF}
=\frac{r^2}{30}+\frac{s(s+2)}{20}-\frac{t^2}{12}.
}
Now we have
\beg{eexam10}{u=(3,3,3)\otimes (2,2,2)=(3,0,0)\otimes (2,1,-1).
}
We begin by choosing the field $c$. Compose first $u$ and
\beg{eexam11}{\overline{u}=(-3,3,-3)\otimes (-2,2,-2)=(-3,0,0)\otimes (-2,1,1).
}
We choose the non-zero $u_n \overline{u}$ of the bottom weight
for the fusion rule which adds the lattice charges on the
right hand side of \rref{eexam10}, \rref{eexam11}. The result
is
\beg{eexam12}{u_{-1/10}\overline{u}=(0,0,0)\otimes (0,2,0).
}
Next, apply $G^{-}_{-1/2}u$ to \rref{eexam12}. Again, we will
choose the bottom descendant. Now $G^{-}_{-1/2}u$ has two
summands,
\beg{eexam13}{(-2,3,1)\otimes (2,1,-1)
}
and
\beg{eexam14}{(3,0,0)\otimes (0,5,3)x_{-1}(-3,1,3)
}
(the term \rref{eexam14} involves renaming to stay withing
no-ghost PF labels after composition). Applying \rref{eexam13}
to \rref{eexam12} gives bottom descendant
\beg{eexam15}{(-2,3,1)\otimes (2,3,-1)\;\text{of weight $8/5$},
}
applying \rref{eexam14} to \rref{eexam12} gives bottom
descendant
\beg{eexam16}{(3,0,0)\otimes (-3,0,0)\;\text{of weight $3/5$}.
}
Since \rref{eexam16} has lower weight than \rref{eexam15},
\rref{eexam15} may be ignored, and we can choose
\beg{eexam17}{c=(3,0,0)\otimes (-3,0,0).
}
Now again, using the formula \rref{eexam6}, we see that
in the sets of fields $G_x$, $G_y$ we need one summand
\rref{eexam14} and three summands \rref{eexam13} to get
to $x_{00}$-charge $0$. Thus, one of the groups $G_x$, 
$G_y$ will contain two summands of \rref{eexam13} and the
other will contain one. We employ the following convention:
\beg{eexam18}{\parbox{3.5in}{We choose $G_y$ to contain
two summands \rref{eexam13} and $G_x$ to contain
one summand \rref{eexam13} and one summand \rref{eexam14}.}
}
This leads to the following:
\beg{eexam18a}{\parbox{3.5in}{We must choose the fields $a$
and $b$ of the same weights and symmetrize the resulting 
correlation function with respect to $x$ and $y$.}
}
We will choose $b$ first. Again, we will choose the bottom
weight (nonzero) descendant of \rref{eexam13} applied to
itself renamed as
\beg{eexam19}{(0,5,-3)x_{-1}(-2,0,-2)\otimes (2,2,2),
}
which is
\beg{eexam20}{(-4,3,-1)\otimes (4,3,1).
}
We rename to level $0$, which gives
\beg{eexam21}{\begin{array}{l}
b=(0,5,3)x_{-1}(-4,0,2)\otimes (0,5,-3)x_{-1}(4,0,-2),\\
w(b)=12/5.
\end{array}
}
Then $a$ must have weight $12/5$ to satisfy \rref{eexam18a}.
When calculating $a$, however, there is an additional
subtlety. This time, we have actually take into account
two summands, from applying \rref{eexam13} to \rref{eexam14}
and vice versa, i.e. \rref{eexam14} to \rref{eexam13}. 
In both cases, we must rename to get the desired fusion
rule. To this end, we may replace \rref{eexam14} by
\beg{eexam22}{(3,0,0)\otimes (-3,2,0).
}
However, when applying \rref{eexam13} and \rref{eexam22} to
each other in opposite order, the renamings then do not correspond,
resulting in the possibility of wrong coefficient/sign (since renaming
are correct only up to constants which we haven't calculated).
To reconcile this, we must use exactly the same renamings
step by step, related only by applying PF currents. To this end,
we may compare the renaming of applying
\beg{eexam23}{(0,5,-3)x_{-1}(-2,0,-2)\otimes (2,2,2)
}
to
\beg{eexam24}{(3,0,0)\otimes \frac{1}{2}(0,5,-3)x_{-1}(-3,1,-3)
}
(the $\frac{1}{2}$ comes from the PF current $(5,-3)x_{-1}(0,-2)$
which takes $(2,2)$ to $2(2,0)$)
and
\beg{eexam25}{(3,0,0)\otimes (-3,2,0)
}
to
\beg{eexam26}{(0,5,-3)x_{-1}(-2,0,-2)\otimes (2,1,-1).
}
We see that the bottom descendant of applying \rref{eexam23} to 
\rref{eexam24} is
\beg{eexam27}{(0,5,-3)x_{-1}(1,0,-2)\otimes(-1)(-1,3,-1)
}
while the bottom descendant of applying \rref{eexam25}
to \rref{eexam26} is
\beg{eexam28}{(0,5,-3)x_{-1}(1,0,-2)\otimes(-1,3,-1).
}
The expression \rref{eexam27} is the negative of \rref{eexam28}.
On the other hand, we see that the bottom descendants of applying
\rref{eexam13} to \rref{eexam22} and vice versa are the same.
This means that we are allowed to use the names \rref{eexam13}
and \rref{eexam22} to each other in either order, but we
must take the results with opposite signs.

Now \rref{eexam28} has weight $7/5$, so to get weight $12/5$,
we must take the descendant of applying \rref{eexam13} to
\rref{eexam22} and vice versa which is of weight $1$ higher than the
bottom. This gives
$$
\begin{array}{l}
((-2,3,1)-(3,0,0))x_{-1}(1,3,1)\otimes (-1,3,-1)+\\
(1,3,1)\otimes ((2,1,-1)-(-3,2,0))x_{-1}(-1,3,-1),
\end{array}
$$
which is
\beg{eexam29}{
a=(-5,3,1)x_{-1}(1,3,1)\otimes (-1,3,-1)+(1,3,1)\otimes (5,-1,-1)x_{-1}(-1,3,-1).
}
Now the correlation function of $a(x)$, $b(y)$, $c(z)$ given in
\rref{eexam29}, \rref{eexam21}, \rref{eexam17} is an ordinary
lattice correlation function. The algorithm for calculating
the lattice correlation function of fields $u_i(x_i)$
which are of the form
$$1_{\lambda_i}(x_i)$$
or 
$$\mu_i x_{-1} 1_{\lambda_i}(x_i)$$
with the label
$$1_{\sum \lambda_i}$$
is as follows: The correlation function is a multiple of
$$\cform{\prod}{i<j}{}(x_i-x_j)^{\langle \lambda_i,\lambda_j\rangle}
$$
by a certain factor, which is a sum over all the ways we may ``absorb"
any $\mu_i x_{-1}$ factors. Each such factor may either be absorbed by
another $\mu_j x_{-1}$, which results in a factor
\beg{eexam30}{\langle\mu_i,\mu_j\rangle (x_i-x_j)^{-2},\; i\neq j
}
or by another lattice label $1_{\lambda_j}$, which results in a factor
\beg{eexam31}{\langle\mu_i,\lambda_j\rangle (x_i-x_j)^{-1},\; i\neq j. 
}
Each $\mu_i x_{-1}$ must be absorbed exactly once (and the mechanism
\rref{eexam30} is considered as absorbing both $\mu_i$ and $\mu_j$),
but one lattice label $1_{\lambda_j}$ may absorb several different
$\mu_i x_{-1}$'s via \rref{eexam31}.

Evaluating the correlation function of $a(x)$, $b(y)$, $c(z)$ with the
vacuum using this algorithm, we get
$$\frac{2(y-z)}{(x-z)(x-y)^3}.$$
Symmetrizing with respect to $x,y$, we get
$$\frac{2(x-2z+y)}{(y-z)(x-z)(x-y)^2},$$
(our total correlation function factor), which is non-zero.

In more detail, we can calculate separately the contributions
to the correlation function of the two summands \rref{eexam29}.
For the first summand, the factor before the $\otimes$ sign
contributes
\beg{eexami1}{-\frac{1}{(x-z)(y-x)},
}
the factor after the $\otimes$ sign contributes
\beg{eexami2}{\frac{1}{y-x}.
}
Multiplying \rref{eexami1} and \rref{eexami2}, we get
$$-\frac{1}{(x-z)(x-y)^2},$$
and symmetrizing with respect to $x$ and $y$,
\beg{eexami3}{-\frac{x-2z+y}{(x-z)(x-y)^2(y-z)},
}
which is the total contribution of the first summand \rref{eexam29}.

For the second summand \rref{eexam29}, the factor after $\otimes$
contributes
\beg{eexami4}{-\frac{2}{(x-y)^2}-\frac{1}{(x-z)(y-x)},
}
and the factor before the $\otimes$ sign contributes
\beg{eexami5}{\frac{1}{y-x}.
}
Multiplying, we get
$$\frac{x-2z+y}{(x-z)(x-y)^3}.$$
After symmetrizing with respect to $x,y$, we get also 
\rref{eexami3}, so both summands of \rref{eexam29} contribute
equally to the correlation function.

\vspace{3mm}
\noindent
{\bf Example 2:}
In this example, we keep the same $a(x)$ and $b(y)$ as in the previous
example, but change $c(z)$. To select $c(z)$, this time we
start with $G^{-}_{-1/2}u$ represented as
\beg{eexam32}{(3,0,0)\otimes (0,5,-3)x_{-1}(-3,1-3)
+C(-2,3,1)\otimes (2,1,-1)
}
($C$ is a non-zero normalization constant which we do not
need to evaluate explicitly), which we apply to $\overline{u}$
represented as
\beg{eexam33}{(-3,0,0)\otimes (-2,1,1).
}
From the two summands \rref{eexam32}, we get bottom descendants
\beg{eexam34}{(0,0,0)\otimes (-5,2,-2)\;\text{of weight $9/10$}
}
and
\beg{eexam35}{(-5,3,1)\otimes (0,2,0)\;\text{of weight $19/10$}.
}
Therefore, we may ignore \rref{eexam35} and select \rref{eexam34}
only. Now applying \rref{eexam34} to $u$ written as
\beg{eexam36}{(3,0,0)\otimes (2,1,-1),
}
we select a descendant of weight $1$ above the label 
$$(3,0,0)\otimes (-3,3,-3).$$
Recalling from the conjugate of \rref{eFst2} that weight $1$ 
states above the label $(3,-3)_{PF}=(0,0)_{PF}$ must vanish, we
get
\beg{eexam37}{c=(3,0,0)\otimes (1,0,0)x_{-1}(-3,0,0)
}
(up to a non-zero multiplicative constant). This gives the correlation
function
\beg{eexam38}{\frac{(x-2z+y)^2}{5(y-z)^2(x-z)^2(x-y)^2}.
}
Let us write again in more detail the contributions
of the two summands \rref{eexam29}. For the first summand,
the contribution of the factor before $\otimes$ is again 
\rref{eexami1} (hasn't changed), and the contribution
of the factor after $\otimes$ is
\beg{eexamj1}{\frac{-y-3z+4x}{15(y-z)(x-z)(x-y)}.
}
Multiplying, we get
$$\frac{-y-3z+4x}{15(y-z)^2(x-z)^2(x-y)},$$
and symmetrizing with respect to $x$, $y$,
\beg{eexamj2}{-\frac{y^2+6yz-6z^2-8xy+6zx+x^2}{15(y-z)^2(x-z)^2(x-y)^2}.
}
This is the total contribution of the first summand 
\rref{eexam29}.

For the second summand \rref{eexam29}, the coordinate before
$\otimes$ contributes again \rref{eexami5}, and the 
coordinate after $\otimes$ contributes
\beg{eexamj3}{\frac{2(-yx-3yz-3xz+3z^2+2x^2+2y^2)}{
15(y-z)(x-z)^2(x-y)^2}.
}
Multiplying, we get
$$-\frac{-yx-3yz-3xz+3z^2+2x^2+2y^2}{15(y-z)(x-z)^2(x-y)^3}.$$
Symmetrizing with respect to $x,y$, we get
\beg{eexamj4}{\frac{2(-yx-3yz-3xz+3z^2+2x^2+2y^2)}{15(y-z)^2(x-z)^2(x-y)^2}
}
which is the total contribution of the second summand
\rref{eexam29}. Adding the contributions \rref{eexamj2}
and \rref{eexamj4} (which are not equal in this case) 
gives \rref{eexam38}.

\vspace{3mm}
The remainder of this Section is dedicated to
comments on possible perturbative deformations along
the fields $(1^5,1^5)$, $(-1^5,1^5)$ (the exponent
here denotes repetition of the field in a tensor
product, and $1$ again stands for $(1,1,1)_{MM}$, etc.). 
We will present some evidence (although not proof) that the obstruction
might vanish in this case. The results we do obtain will prove
useful in the next Section.
Such conjecture would have a geometric interpretation.
In Gepner's conjectured interpretation of the model we are investigating
as the $\sigma$-model of the Fermat quintic, the field \rref{eg20}
corresponds to the dilaton. It seems reasonable to conjecture that
the dilaton deformation should exist, since the theory should not
choose a particular global size of the quintic. Similarly, the field
\rref{eg19} can be explained as the dilaton on the mirror manifold
of the quintic, which should correspond to deformations of complex
structure of the form
\beg{eg65}{x^5+y^5+z^5+t^5+u^5+\lambda xyztu=0.
}
Therefore, our analysis predicts that the (body of) the
moduli space of $N=2$-supersymmetric CFT's containing the Gepner
model is $2$-dimensional, and contains $\sigma$-models of the
quintics \rref{eg65}, where the metric is any multiple of the
metric for which the $\sigma$-model exists (which is unique up
to a scalar multiple).

\vspace{3mm}
To discuss possible deformations along the fields
$(1^5,1^5)$, $(-1^5,1^5)$, let us first
review a simpler case, namely the
coset construction: In a VOA $V$, we set, for $u\in V$
homogeneous,
\beg{eg66}{\begin{array}{l}
Y(u,z)=\cform{\sum}{n\in\Z}{}u_{-n+w(u)}z^n,\\
Y_{-}(u,z)=\cform{\sum}{n<0}{}u_{-n+w(u)}z^n,\\
Y_{+}(u,z)=Y(u,z)-Y_{-}(u,z).
\end{array}
}
The coset model of $u$ is
\beg{eg67}{\begin{array}{l}V_{u}=\\
\langle v\in V|Y_{-}(u,z)v=0\;
\text{and $Y_{+}(u,z)$ involves only integral powers of $z$}
\rangle.
\end{array}
}
Then $V_u$ is a sub-VOA of $V$. To see this, recall that
\beg{eg68}{
Y(u,z)Y(v,t)w=\\
Y_{+}(u,z)Y_{-}(v,t)w+Y_{+}(v,t)Y_{-}(u,z)w+Y(Y_{-}(u,z-t)v)w.
}
When $v,w\in V_u$, the last two terms of the right hand side of \rref{eg68}
vanish, which proves that 
$$Y(v,t)w\in V_u[[t]][t^{-1}].$$
Now in the case of $N=2$-super-VOA's, let us stick to the NS sector.
Then \rref{eg66} still correctly describes the ``body'' of a vertex
operator. The complete vertex operator takes the form
\beg{eg69}{\begin{array}{l}
Y(u,z,\theta^+,\theta^-)=\\
\cform{\sum}{n\in\Z}{}
u_{-n+w(u)}z^n+u_{-n-1/2+w(u)}^{+}z^n\theta^+
+u_{-n-1/2+w(u)}^{-}z^n\theta^-
+u_{-n-1+w(u)}^{\pm}z^n\theta^+\theta^-.
\end{array}
}
We still define $Y_{-}(u,z,\theta^+,\theta^-)$ to be the sum of terms
involving $n<0$, and $Y_+(u,z,\theta^+,\theta^-)$ the sum of the remaining
terms.
The compatibility relations for an $N=2$-super-VOA are
\beg{eg70}{\begin{array}{l}
D^+Y(u,z,\theta^+,\theta^-)=Y(G^{+}_{-1/2}u,z,\theta^+,\theta^-)\\
D^-Y(u,z,\theta^+,\theta^-)=Y(G^{-}_{-1/2}u,z,\theta^+,\theta^-)
\end{array}
}
where
\beg{eg71}{D^+=\frac{\partial}{\partial\theta^+}+\theta^+\frac{\partial}{
\partial z},\;\; 
D^-=\frac{\partial}{\partial\theta^-}+\theta^-\frac{\partial}{
\partial z}.
}
Now using \rref{eg68} again, for $u\in V$ homogeneous, we will have
a sub-$N=2$-VOA $V_u$ defined by \rref{eg67}, which is further endowed
with the operators $G^{-}_{-1/2}$, $G^{+}_{-1/2}$.

\vspace{3mm}
In the case of lack of locality, only a weaker conclusion holds.
Assume first we have ``abelian'' fusion rules in the
same sense as in Remark 2 after Theorem \ref{l1}.


\begin{lemma}
\label{l2}
Suppose we have fields $u_i, i=0,...,n$ such that for $i>j$,
\beg{ecos1}{Y(u_i,z)u_j=\cform{\sum}{n\geq 0}{}(u_i)_{-n-\alpha_{ij}-w(u_i)}
z^{n+\alpha_{ij}}
}
with $0\leq \alpha_{ij}<1$. Consider further points $z_0=0$, $z_1,...,z_n$.
Then 
\beg{ecos2}{\cform{\prod}{n\geq i>j\geq 0}{}
(z_i-z_j)^{-\alpha_{ij}}Y(u_n,z_n)...Y(u_z,z_1)u_0
}
where each $(z_i-z_j)^{-\alpha_{ij}}$ are expanded in $z_j$
is a power series whose coefficients involve nonnegative integral
powers of $z_0,...,z_n$ only.
\end{lemma}

\vspace{3mm}
\Proof
Induction on $n$. Assuming the statement is true for $n-1$, note that
by assumption, \rref{ecos2}, when coupled to $w^{\prime}\in V^{\vee}$
of finite weight, is a meromorphic function in $z_n$ with possible
singularities at $z_0=0,z_1,...,z_n-1$. Thus, \rref{ecos2} can be
expanded at its singularities, and is equal to
\beg{ecos3}{
\begin{array}{l}
\cform{\prod}{n-1\geq i>j\geq 0}{}(z_i-z_j)^{-\alpha_{ij}}\cdot\\
( (z_{n}^{-\alpha_{n0}}expand_{z_n}\cform{\prod}{j\neq 0}{}(z_n-z_j)^{-\alpha_{nj}}
Y(u_{n-1},z_{n-1})...Y(u_1,z_1)Y(u_n,z_n)u_0)_{z_{n}^{<0}}\\
+(\cform{\sum}{i=1}{n-1}(z_n-z_i)^{-\alpha_{ni}}
expand_{(z_n-z_i)}(\cform{\prod}{n-1\geq j\neq i}{}(z_n-z_j)^{-\alpha_{nj}})\cdot\\
Y(u_{n-1},z_{n-1})...Y(u_{i+1},z_{i+1})Y(Y(u_n,z_n-z_i)u_i,z_i)\cdot\\
Y(u_{i-1},z_{i-1})...Y(u_1,z_1)u_0)_{(z_n-z_i)^{<0}}\\
+(z_{n}^{-\alpha_{n0}-...-\alpha_{n,n-1}}expand_{1/z_n}\cform{\prod}{j=1}{n-1}
(1-\frac{z_j}{z_n})^{-\alpha_{nj}}\cdot\\
Y(u_n,z_n)Y(u_{n-1},z_{n-1})...Y(u_1,z_1)u_0)_{z_{n}^{\geq 0}}).
\end{array}
}
In \rref{ecos3}, $expand_{?}(?)$ means that the argument is expanded in the
variable given as the subscript. The symbol $(?)_{?^{<0}}$ (resp.
$(?)_{?^{\geq 0}}$) means that we take only terms in the argument,
(which is a power series in the subscript), which involve negative
(resp. non-negative) powers of the subscript. 

In any case, by the assumption of the Lemma, all summands \rref{ecos3} vanish
with the exception of the last, which is the induction step.
\qed

\vspace{3mm}
In the case of non-abelian fusion rules, an analogous
result unfortunately fails.
Assume for simplicity that
\beg{efff+}{\parbox{3.5in}{$u_0=...=u_n$ holds in \rref{ecos3}
with $0\leq \alpha_{ij}^{F}<1$ true for any fusion rule $F$.}} 
We would like to conclude that 
the correlation function
\beg{e10l1}{\langle v, u(z_n)...u(z_1)u\rangle
}
involves only non-negative powes of $z_i$ when expanded in
$z_1,...,z_n$ (in this order). Unfortunately, this is
not necessarily the case. Note that we know that \rref{e10l1}
converges to $0$ when two of the argument $z_i$
approach while the others remain separate. However, this
does not imply that the function \rref{e10l1} converges to $0$
when three or more of the arguments approach simultaneously.

To give an example, let us consider the solution of the Fuchsian differential
equation of $\P^1-\{0,t,\infty\}$
\beg{efff*}{y^\prime=(\frac{A}{x}+\frac{B}{z-t})y
}
for square matrices $A,B$ (with $t\neq 0$ constant). Since
the solution $y$ has bounded singularities, multiplying $y$
by $z^m(z-t)^n$ for large enough integers $m$, $n$ makes the resulting
function $Y$
converge to $0$ when $z$ approaches $0$ or $t$. If, however, the
expansion of $Y$ at $\infty$ involved only non-negative powers of $z$,
it would have only finitely many terms, and hence abelian monodromy. 
It is well known, however, that this is not necessarily the case. In
fact, any irreducible monodromy occurs for a solution of the equation \rref{efff*}
for suitable matrices $A$, $B$ (cf. \cite{anbo}).

\vspace{3mm}
Therefore, the following result may be used as evidence, but
not proof, of the exponentiability of deformations along $(1^5, 1^5)$
and $(1^5,-1^5)$.

\begin{lemma}
\label{l20}
The assumption \rref{efff+} is satisfied for the
field
$$u=G^{-}_{-1/2}((1,1,1),...,(1,1,1))
$$
in the $5$-fold tensor product of the $N=2$ minimal model
of central charge $9/5$.
\end{lemma}

\vspace{3mm}
Before proving this, let us state the following consequence:

Indeed, assuming Lemma \ref{l20} and setting $w=((1,1,1),...,(1,1,1))$,
the obstruction is
\beg{e10t2}{\langle w^{\prime}| (G^{-}_{-1/2})w(z_n)...
(G^{-}_{-1/2}w)(z_1)w\rangle.
}
(The antichiral primary case is analogous.) But using
the fact that
$$G^{-}_{-1/2}((G^{-}_{-1/2})w(z_n)...
(G^{-}_{-1/2}w)(z_1)w)=(G^{-}_{-1/2})w(z_n)...
(G^{-}_{-1/2}w)(z_1)G^{-}_{-1/2}w
$$
along with injectivity of $G^{-}_{-1/2}$ on chiral
primaries of weight $1/2$, we see that the non-vanishing
of \rref{e10t2} implies the non-vanishing of \rref{e10l1}
with $u=G^{-}_{-1/2}w$ for some $v$ of weight $1$, which
would contradict Lemma \rref{l20}.

\vspace{3mm}
\noindent
{\bf Proof of Lemma \ref{l20}:}
We have 
\beg{e20l3a}{G^{-}_{-1/2}(1,1,1)=(-4,1,-1).
}
We have in our lattice
\beg{e20l3}{\begin{array}{l}
(1,1,1)\cdot(1,1,1)=1/15+1/10-1/6=0,\\
(-4,1,-1)\cdot(-4,1,-1)=16/15+1/10-1/6=1,\\
(1,1,1)\cdot (-4,1,-1)=-4/15+1/10+1/6=0,
\end{array}
}
so we see that with the fusion rules which stays on levels $1,2$,
the vertex operators $u(z)u$ have only non-singular terms.

However, this is not sufficient to verify \rref{efff+}. In effect,
when we use the fusion rule which goes to levels $0,3$,
$$(1,1,1)(z)(-4,1,-1)$$ 
and 
$$(-4,1,-1)(z)(1,1,1)$$ 
will have most singular term $z^{-2/5}$, so when we write
again $1$ instead of $(1,1,1)$ and $G$ instead of $G^{-}_{-1/2}(1,1,1)$,
with the least favorable choice of fusion rules, 
it seems $u(z)u$ can have singular term $z^{-4/5}$, coming from
the expressions
\beg{e20l4}{(G1111)(z)(1G111)
}
and
\beg{e20l5}{(1G111)(z)(G1111)
}
(and expression obtained by permuting coordinates). Note that with
other combinations of fusion rules, various other singular terms
can arise with $z^{>-4/5}$.

Now the point is, however, that we will show that with any choice
of fusion rule, the most singular terms of \rref{e20l4} and
\rref{e20l5} come with opposite signs and hence cancel out.
Since the $z$ exponents of other terms are higher by an integer,
this is all we need.

Recalling the Koszul-Milnor sign rules for the minimal model,
recall that $1$ is odd and $G$ is even, so
\beg{e20l6}{(G\otimes 1)(z)(1\otimes G)=-G(z)1\otimes 1(z)G,
}
\beg{e20l7}{(1\otimes G)(z)(G\otimes 1)=1(z)G\otimes G(z)1.
}
We have
$$1(z)G=(1,1,1)(z)(-4,2,2)=M(-3,0,0)z^{-2/5}+\text{HOT},$$
$$G(Z)1=(-4,2,2)(z)(1,1,1)=N(-3,0,0)z^{-2/5}+\text{HOT},$$
with some non-zero coefficients $M,N$, so the bottom descendants of
\rref{e20l6} and \rref{e20l7} are
$$-MN(-3,0,0)\otimes(-3,0,0)$$
resp.
$$MN(-3,0,0)\otimes(-3,0,0),$$
so they cancel out, as required.
\qed

\vspace{5mm}

\section{The case of the Fermat quartic $K3$-surface}
\label{sk3}

The Gepner model of the $K3$ Fermat quartic
is an orbifold analogous to
\rref{egepp2} with $5$ replaced by $4$ of the $4$-fold
tensor product of the level $2$ $N=2$-minimal model, although
one must be careful about certain subtlteties arising
from the fact that the level is even. The model has central
charge $6$. The level $2$ PF model is the $1$-dimensional
fermion (of central charge $1/2$), viewed
as a bosonic CFT. As such, that model has $3$ labels,
the NS label with integral weights (denote by $NS$), the NS label with 
weights $\Z+\frac{1}{2}$ (denote by $NS^{\prime}$), and the
R label (denote by $R$). The fusion rules are given by 
the fact that $NS$ is the unit label,
\beg{ek31}{\begin{array}{l}NS^{\prime}*NS^{\prime}=NS,\\
NS^{\prime} *R=R,\\
R*R=NS +NS^{\prime}.
\end{array}
}
We shall again find it useful to use the free field 
realization of the $N=2$ minimal model, 
which we used in the last two sections. 
In the present case, the theory is a subquotient
of a lattice theory spanned by
$$(\frac{1}{\sqrt{2}},0,0),\; (\frac{1}{\sqrt{8}},\frac{i}{\sqrt{8}},
\frac{i}{2}),\;(\frac{1}{\sqrt{2}},0,i).$$
Analogously as before, we write $(k,\ell,m)$
for
$$(\frac{k}{\sqrt{8}},\frac{\ell i}{\sqrt{8}},\frac{mi}{2}).$$
The conformal vector is
$$\frac{1}{2}x_{0,-1}^{2}-\frac{1}{2}x_{1,-1}^{2}+\frac{i}{2\sqrt{2}}
x_{1,-2}+\frac{1}{2}x_{2,-1}^{2}.$$
The superconformal vectors are
$$G_{-3/2}^{-}=(0,4,2)x_{-1}(4,0,2),$$
$$G_{-3/2}^{+}=(0,4,-2)x_{-1}(-4,0,-2).$$
The fermionic labels will again be denoted by omitting
the first coordinate: $(\ell,m)_F$. The fermionic identifications
are:
\beg{ek3+}{\begin{array}{l}(2,2)_F\sim (2,-2)_F \sim (0,0)_F\\
(1,1)_F\sim (1,-1)_F.
\end{array}
}
A priori the lattice $\langle\sqrt{8}\rangle$ has $8$ labels
$\frac{k}{\sqrt{8}}$, $0\leq k\leq 7$, but the $G^-$
definition together with \rref{ek3+} forces the MM identification
of labels
$$(1,1,1)\sim (-3,1,-1)\sim (-3,1,1).$$
The labels of the level $2$ MM are therefore
$$\begin{array}{l}
(2k,0,0),\;0\leq k\leq 3,\\
(2k+1,1,1),\; 0\leq k\leq 1.
\end{array}
$$
The fusion rules are
$$\begin{array}{l}
(k,0,0)*(\ell,0,0)=(k+\ell,0,0),\\
(k,0,0)*(\ell,1,1)=(k+\ell,1,1),\\
(k,1,1)*(\ell,1,1)=(k+\ell,0,0)+(k+\ell+4,0,0),
\end{array}
$$
so the Verlinde algebra is simply
$$\Z[a,b]/(a^4=1,\;b^2=a+a^3,\;a^2b=b)$$
where $a=(2,0,0)$, $b=(1,1,1)$.

One subtlety of the even level MM in comparison with odd
level concerns signs. Since the $k$-coordinates of $G^-$ and
$G^+$ are even, we can no longer use the $k$-coordinate of
an element as an indication of parity ($u$ and $G^{\pm}u$ cannot
have the same parity). Because of this, we must introduce
odd fusion rules. There are various ways of doing this.
For example, let the bottom states of
$(2k,0,0)$, $(1,1,1)$ and $(-1,1,1)$ be even. Then the
fusion rules on level $\ell=0$ are even, as are the
fusion rules combining levels $0$ and $1$.
The fusion rules
$$(1,1,1)*(1,1,1)\mapsto (2,0,0), \; (1,1,1)*(-1,1,1)\mapsto (2,0,0)$$
are even, the remaining fusion rules (adding $4$ to the $k$-coordinate
on the right hand side) are odd. 

\vspace{3mm}
Now the $c$ fields of the MM are
$$(0,0,0),\; (1,1,1), \; (2,2,2)$$
and the $a$ fields are
$$(0,0,0), \; (-1,1,-1), \; (-2,2,-2).$$
If we denote by $H_{1,2k+1}$ the state space of label $(2k+1,1,1)$,
$0\leq k\leq 1$,
and by $H_{0,2k}$ the state space of label $(2k,0,0)$, $0\leq k \leq 3$,
them the state space of the $4$-fold tensor product of the
level $2$ minimal model is 
\beg{ek3p1}{\cform{\hat{\bigotimes}}{i=0}{3}((\cform{\bigoplus}{k_i=0}{3}
H_{0,2k_i}\hat{\otimes}H_{0,2k_i}^{*})\oplus 
(\cform{\bigoplus}{k_i=0}{1}
H_{1,2k_i+1}\hat{\otimes}H_{1,2k_i+1}^{*})).
}
The Gepner model is an orbifold with respect to the $\Z/4$-group which
acts by $i^\ell$ on products in \rref{ek3p1} where the sum of
the subscripts $2k_i$ or $2k_i+1$ is congruent to $\ell$ modulo $4$.
Therefore, the state space of the Gepner model is the sum over $\beta\in\Z/4$
and $\alpha_i\in\Z/4$, 
$$\cform{\sum}{i=0}{3}\alpha_i=0\in\Z/4,$$
of
\beg{ek3p2}{\cform{\hat{\bigotimes}}{i=0}{3}((\cform{\bigoplus}{2k_i\equiv \alpha_i\mod
4}{}
H_{0,2k_i}\hat{\otimes}H_{0,2k_i+2\beta}^{*})\oplus 
(\cform{\bigoplus}{2k_i+1\equiv\alpha_i}{}
H_{1,2k_i+1}\hat{\otimes}H_{1,2k_i+1+2\beta}^{*})).
}
It is important to note that each summand \rref{ek3p2}
in which all the factors have 
the ``odd'' subscripts $1,2k_i+1$ occurs twice in the orbifold state space. 

If we write again $\ell$ for $(\ell,\ell,\ell)$ and $-\ell$
for $(-\ell,\ell,-\ell)$, then the critical $cc$ fields are chirally
symmetric permutations of
\beg{ek3p3}{\begin{array}{l}
(2,2,0,0),\;(2,2,0,0)\\
(2,1,1,0),\;(2,1,1,0)\\
(1,1,1,1),\;(1,1,1,1).
\end{array}
}
Note that applying all the possible permutations to the
fields \rref{ek3p3}, we obtain only $19$ fields, while there should
be $20$, which is the rank of $H^{11}(X)$ for a $K3$-surface
$X$. However, this is where the preceeding comment comes to play:
the last field \rref{ek3p3} corresponds to a term \rref{ek3p2}
where all the factors have odd subscripts, and hence there 
are two copies of that summand in the model, so the last
field \rref{ek3p3} occurs ``twice''. 

By the fact that the Fermat quartic Gepner model has
$N=(4,4)$ worldsheet supersymmetry (se e.g. \cite{am,nw} and references
therein), the spectral flow guarantees that the number of critical
$ac$ fields
is the same as the number of critical $cc$ fields. Concretely, the
critical $ac$ fields are the permutations of
\beg{ek3p4}{\begin{array}{l}
(0,0,-2,-2),\;(2,2,0,0)\\
(0,-1,-1,-2),\;(2,1,1,0)\\
(-1,-1,-1,-1),\;(1,1,1,1).
\end{array}
}
As above, the last field \rref{ek3p4} occurs in $2$ copies,
thus the rank of the space of critical $ac$ fields is also $20$.  

\vspace{3mm}

We wish to investigate whether infinitesimal deformations
along the fields \rref{ek3p3}, \rref{ek3p4} exponentiate
perturbatively. To this end, let us first see when the
``coset-type scenario'' occurs. This is sufficient
to prove convergence in the present case. This is due to the
fact that in the present theory, there is an even number of
fermions, in which case it is well known by the
boson-fermion correspondence that the correlation functions
follow abelian fusion rules, and therefore Lemma \ref{l2} applies.
To prove that the coset scenario occurs, let us look
at the chiral $c$ fields 
$$u=(2,2,0,0), \;(2,1,1,0), \; (1,1,1,1)
$$
and study the singularities of
\beg{ek3q2a}{G^{-}_{-1/2}(z)(G^{-}_{-1/2}u).
}
By Lemma \ref{l2},
if \rref{ek3q2a} are non-singular, the obstructions vanish.
The inner product is
$$
\begin{array}{l}
\langle(k,\ell,m),(k^{\prime},\ell^{\prime},m^{\prime})\rangle
=\frac{kk^{\prime}}{8}+\frac{\ell\ell^{\prime}}{8}-\frac{mm^{\prime}}{4},
\\
w(k,\ell,m)=\frac{k^2}{16}+\frac{\ell(\ell+2)}{16}-\frac{m^2}{8}.
\end{array}
$$
Next, $(2,2,2)=(2,0,0)$,
$$G^{-}_{-1/2}(2,0,0)=(0,4,-2)x_{-1}(-2,0,-2),
$$
$$G^{-}_{-1/2}(1,1,1)=(-3,1,-1).
$$
if we again replace, to simplify notation, the symbol $G^{-}_{-1/2}$
by $G$, then we have
\beg{ek3q3}{\parbox{3.5in}{The most singular $z$-power of $2(z)2$
is $\frac{2\cdot 2}{8}=\frac{1}{2}$,}
}
\beg{ek3q4}{\parbox{3.5in}{The most singular $z$-power of $G2(z)2$
is $\frac{-2\cdot 2}{8}=-\frac{1}{2}$.}
}
For $G2(z)G2$, rename the rightmost $G2$ as $(-2,2,0)$. We get
\beg{ek3q5}{\parbox{3.5in}{The most singular $z$-power of $G2(z)G2$
is $-1+\frac{(-2)\cdot (-2)}{8}=-\frac{1}{2}$,}
}
\beg{ek3q6}{\parbox{3.5in}{The most singular $z$-power of $1(z)1$
is $0$ for the even fusion rule and $1/2$ for the odd fusion rule,}
}
\beg{ek3q7}{\parbox{3.5in}{The most singular $z$-power of $G1(z)1$
is $\frac{-3}{8}+\frac{1}{8}+\frac{1}{4}=0$ for the even fusion rule, 
and $-1/2$ for the odd fusion rule,}
}
\beg{ek3q8}{\parbox{3.5in}{The most singular $z$-power of $G1(z)G1$
is $\frac{9}{8}+\frac{1}{8}-\frac{1}{4}=1$ 
for the even fusion rule and $3/2$ for the odd fusion rule.}
}
One therefore sees that for the field $u=(1,1,1,1)$,
\rref{ek3q2a} is non-singular: In the case of the least
favorable (odd) fusion rules, the most singular term
appears to be $-1$, coming from 
\beg{ek3qa}{(G1,1,1,1)\otimes(1,G1,1,1).
}
However, this term cancels with
\beg{ek3qb}{(1,G1,1,1)\otimes(G1,1,1,1).
}
To see this, note that the last two coordinates do not enter
the picture. We have an odd (resp. even) pair of pants
$P_-$ resp. $P_+$ in the MM with input $1,1$. They add up
to a pair of pants in MM$\otimes$MM. On \rref{ek3qa},
we have pairs of pants $P_i\in\{P_-,P_+\}$, 
\beg{ek3qc}{\begin{array}{l}
P(G1\otimes1)\otimes(1\otimes G1)=\\
(P_1\otimes P_2)(G1\otimes1)\otimes(1\otimes G1)=\\
sP_1(G1\otimes 1)\otimes P_2(1\otimes G1)
\end{array}
}
where $s$ is the sign of permuting $P_2$ past $G1\otimes 1$.
Here we use the fact that $1$ is even. On the other hand,
\beg{ek3qd}{\begin{array}{l}
P(1\otimes G1)\otimes(G1\otimes 1)=\\
(P_1\otimes P_2)(1\otimes G1)\otimes(G1\otimes 1)=\\
-sP_1(1\otimes G1)\otimes P_2(G1\otimes 1)
\end{array}
}
(as $G1$ is odd, so there is a $-$ by permuting it with itself).
From \rref{ek3q*}, the lowest term of $P_i(1\otimes G1)$
and $P_i(G1\otimes 1)$ have opposite signs, so \rref{ek3qc}
and \rref{ek3qd} cancel out. 

\vspace{3mm}
The situation is simpler for $u=(2,2,0,0)$, in which case all 
the fusion rules are even, and the most singular term of
$$(G2\otimes 2)(z)(2\otimes G2)$$
appears to have most singular term $z^{-1}$. However, again
note that $2$ is even and $G2$ is odd, so
\beg{ek3q30}{(G2\otimes s)(z)(2\otimes G2)=G2(z)2\otimes 2(z)G2,
}
while
\beg{ek3q40}{(2\otimes G2)(z)(G2\otimes 2)=-2(z)G(z)\otimes
G2(z)2.
}
Renaming $G2$ as $(-2,2,0)$, the bottom descendant of both
$G2(z)2$ and $2(z)G2$ is $(0,2,0)$ with some coefficient, so 
\rref{ek3q30} and \rref{ek3q40} cancel out. Thus, the 
deformations along the first and
last fields of \rref{ek3p3} and \rref{ek3p4} exponentiate.

\vspace{3mm}
The field $u=(2,1,1,0)$ is difficult to analyse, since 
in this case, \rref{ek3q2a} has singular channels and
the coset-type scenario does not occur. We do not know
how to calculate the obstruction directly in this
case. It is however possible to present an indirect
argument why these deformations exist. 

In one precise formulation, the boson-fermion correspondence
asserts that a tensor product of two copies of the 
$1$-dimensional chiral fermion theory considered
bosonically (= the level $2$ parafermion) is an
orbifold of the lattice theory $\langle 2\rangle$,
by the $\Z/2$-group whose generator acts on the lattice
by sign. 

This has an $N=2$-supersymmetric version. 
We tensor with two copies of the lattice theory associated with 
$\langle\sqrt{8}\rangle$, picking out the sector
\beg{e21101}{(\frac{m}{\sqrt{8}},\frac{n}{\sqrt{8}},\frac{p}{4})\;
\text{where $m\equiv n\equiv p \mod 2$.}
}
The fermionic currents of the individual coordinates are
\beg{e21102}{\psi_{-1/2,1}=(1)+(-1),\; \psi_{-1/2,2}=i((1)-(-1)),
}
so the SUSY generators are
\beg{e21103}{G^{\pm}_{-3/2,1}=(\pm\frac{4}{\sqrt{8}},0)\otimes((1)+(-1)),
\; G^{\pm}_{-3/2,2}=(0,\pm\frac{4}{\sqrt{8}})\otimes i((1)+(-1)),
}
$$G=G_{\cdot 1}+G_{\cdot 2}.
$$
The $\Z/2$ group acts trivially on the new lattice coordinate.

A note is due on the signs: To each state, we can assign
a pair of parities, which will correspond to the parities
of the $2$ coordinates in the orbifold. This then also
determines the sign of fusion rules. 

Now consider our field as a tensor product of $(2,0)$ and
$(1,1)$, each in a tensor product of two copies of the minimal
models. Considering each of these factors as orbifold of
the $N=2$-supersymmetric lattice theory, let us lift to
the lattice theory:
\beg{e21104}{(2,0)\mapsto (\frac{2}{\sqrt{8}},0)\otimes (0),
}
\beg{e21105}{(1,1)\mapsto (\frac{1}{\sqrt{8}},\frac{1}{\sqrt{8}})
\otimes ((1/2)+(-1/2)).
}
Then the fields \rref{e21104}, \rref{e21105} are $\Z/2$-invariant.
In the case of \rref{e21104}, we can proceed in the lift instead of
the orbifold, because the fusion rules in the orbifold are
abelian anyway. In the case of \rref{e21105}, the choice amounts
to choosing a particular fusion rule. But now the point is
that
\beg{e21106}{G^{-}_{-1/2}(2,0)\mapsto (-\frac{2}{\sqrt{8}},0)
\otimes((1)+(-1)),
}
\beg{e21107}{G^{-}_{-1/2}(1,1)\mapsto (-\frac{3}{\sqrt{8}},\frac{1}{\sqrt{8}})
\otimes((1/2)+(-1/2)),
}
\beg{e21108}{G^{-}_{-1/2}(1,2)\mapsto (\frac{1}{\sqrt{8}},-\frac{3}{\sqrt{8}})
\otimes((1/2)+(-1/2)).
}
Thus, the left of $G^{-}_{-1/2}u$ is a sum of lattice labels!

Now the critical summands of the operator
\beg{e21109}{G^{-}_{-1/2}(u)(z_k)...G^{-}_{-1/2}(u)(z_1)(u)(0)
}
have $k=4m$, and we have $2m$ summands \rref{e21106},
and $m$ summands \rref{e21107}, \rref{e21108}, respectively. 
All
$$\left(\begin{array}{c}4m\\2m,m,m
\end{array}
\right)$$
possibilities occur. It is the bottom (=label) term which we
must compute in order to evaluate our obstruction. But by
our sign discussion, when we swap a \rref{e21107} term with
a \rref{e21108} term, the label summands cancel
out. Now adding all such possible 
$$\left(\begin{array}{c}4m\\2m,m-1,m-1,2
\end{array}
\right)$$
pairs, all critical summands of \rref{e21109} will occur
with equal coefficients by symmetry, and hence also the
bottom coefficient of \rref{e21102} is $0$, thus showing
the vanishing of our obstruction for this field lift to
the lattice theory. 

Since the field \rref{e21105} is invariant under the 
$\Z/2$-orbifolding (and although \rref{e21104} isn't, 
the same conclusion holds when replacing it with its orbifold
image), the entire perturbative
deformation can also be orbifolded, yielding the
desired deformation.

\vspace{3mm}
We thus conclude that for the Gepner model of the
$K3$ Fermat quartic, all the critical fields exponentiate
to perturbative deformations.

\vspace{5mm}

\section{Conclusions and discussion}
\label{scd}

In this paper, we have investigated perturbative deformations
of CFT's by turning on a marginal $cc$ field, by the 
method of recursively updating the field along the
deformation path. A certain algebraic obstruction arises.
We work out some examples, including free field theories,
and some $N=(2,2)$ supersymmetric Gepner models. In the
$N=(2,2)$ case, in the case of a single $cc$ field, the 
obstruction we find can be made very explicit, and perhaps
surprisingly, does not automatically vanish. By explicit
computation, we found that the obstruction does not
vanish for a particular critical $cc$ field in the
Gepner model of the Fermat quintic $3$-fold (we saw
some indication, although not proof, that
it may vanish for the field corresponding to adding the
symmetric term $xyztu$ to the superpotential, and for
the unique critical $ac$ field). By comparison, the obstruction
vanishes for the critical $cc$ fields and
$ac$ fields Gepner model of the Fermat quartic $K3$-surface.
Our calculations are not completely physical in the sense
that $cc$ fields are not real: real fields
are obtained by adding in each case the complex-conjugate 
$aa$ field, in which case the calculation is more complicated
and is not done here. 

\vspace{3mm}
Assuming (as seems likely) that the real field case exhibits
similar behaviour as we found, why are the $K3$ and $3$-fold
cases different, and what does the obstruction in the 
$3$-fold case indicate? In the $K3$-case, our perturbative
analysis conforms with the Aspinwall-Morrison \cite{am}
of the big moduli space of $K3$'s, and corresponding $(2,2)$-
(in fact, $(4,4)$-) CFT's, and also with the findings of
Nahm and Wendland \cite{nw,wend}.

\vspace{3mm}
In the $3$-fold case, however, the straightforward perturbative
construction of the deformed non-linear $\sigma$-model
fails. This corresponds to the discussion of Nemeschansky-Sen
\cite{ns} of the renormalization of the non-linear $\sigma$-model.
They expand around the $0$ curvature tensor, but it seems
natural to assume that similar phenomena would occur if we
could expand around the Fermat quintic vacuum. Then 
\cite{ns} find that non-Ricci flat deformations must be added
to the Lagrangian at higher orders of the deformation parameter
in order to cancel the $\beta$ function. Therefore, if we
want to do this perturbatively, fields must be present in
the original (unperturbed) model which would correspond to
non-Ricci flat deformation. No such fields are present
in the Gepner model. (Even if we do not a priori
assume that the marginal fields of the Gepner model
correspond to Ricci flat deformations, we see that
{\em different} fields are needed at higher order of
the perturbation parameter, so there are not enough fields 
in the model.) More generally, ignoring for the
moment the worldsheet SUSY, the bosonic
superpartners are fields which are of weight $1$ classically
(as the classical non-linear $\sigma$-model Lagrangian is
conformally invariant even in the non-Ricci flat case).
A $1$-loop correction arises in the quantum picture
\cite{af}, indicating
that the corresponding deformation fields must be of generalized weight
(cf. \cite{zhang,zhang1,huang,huang1}). However, such fields are excluded in
unitary CFT's, which is the reason why these deformations
must be non-perturbative. One does not see this phenomenon
on the level of the corresponding topological models, since
these are invariant under varying the metric within the
same cohomological class, and hence do not see the correction
term \cite{topsym}. Also, it is worth noting that
in the $K3$-case, the $\beta$ function vanishes directly
for the Ricci-flat metric by the $N=(4,4)$ 
supersymmetry (\cite{agg}), and hence the correction terms of \cite{ns}
are not needed. Accordingly, we have found that the corresponding
perturbative deformations exist. 

\vspace{3mm}
From the point of view of mirror symmetry, mirror-symmetric
families of hypersurfaces in toric varieties were proposed
by Batyrev \cite{bat}. In the case of the Fermat quintic,
the exact mirror is a singular orbifold and the non-linear
$\sigma$-model deformations corresponding to the Batyrev
dual family exist perturbatively by our analysis. To obtain
mirror candidates for the additional deformations, one uses
crepant resolutions of the mirror orbifold (see \cite{reid}
for a survey). In the $K3$-case, this approach seems validated
by the fact that the mirror orbifolds can indeed be viewed
as a limit of non-singular $K3$-surfaces \cite{and}. In 
the $3$-fold case, however, this is not so clear. The 
moduli spaces of Calabi-Yau $3$-folds are not locally symmetric
spaces. The crepant resolution is not unique even in the
more restrictive category of algebraic varieties; different
resolutions are merely related by flops. It is therefore
not clear what the exact mirrors are of those deformations of
the Fermat quntic where the deformation does not naturally
occur in the Batyrev family, and resolution of singularities
is needed. In other words, the McKay correspondence
sees only ``topological'' invariants, and not the finer 
geometrical information
present in the whole non-linear $\sigma$ model.

\vspace{3mm}
In \cite{ruan}, Fan, Jarvis and Ruan constructed exactly mathematically
the $A$-models corresponding to Landau-Ginzburg orbifolds via
Gromov-Witten theory applied to the Witten equation. Using
mirror symmetry conjectures, this may be used to construct
mathematically candidates of topological gravity-coupled $A$-models
as well as $B$-models of Calabi-Yau varieties. Gromov-Witten theory,
however, is a rich source of examples where such gravity-coupled
topological models exist, while a full conformally invariant 
$(2,2)$-$\sigma$-model does not. For example, Gromov-Witten theory
can produce highly non-trivial topological models for $0$-dimensional
orbifolds (cf. \cite{okpa,pdjo}).

\vspace{3mm}
Why does our analysis not contradict the calculation
of Dixon \cite{dixon} that the central charge does not change
for deformation of any $N=2$ CFT along any linear combination
of $ac$ and $cc$ fields? Zamolodchikov \cite{za,zb} defined
an invariant $c$ which is a non-decreasing function in
a renormalization group flow in a $2$-dimensional QFT,
and is equal to the central charge in a conformal field
theory. It may therefore appear that by \cite{dixon},
all infinitesimal deformations along critical $ac$ and
$cc$ fields in an $N=2$-CFT exponentiate. However, we saw
that when our obstruction occurs, additional counterterms
corresponding to those of Nemeschansky-Sen are needed.
This corresponds to non-perturbative corrections of the
correlation function needed to fix $c$, and the functions \cite{dixon}
cannot be used directly in our case. 

\vfill
\pagebreak

\end{document}